\documentclass[pra,aps,amsmath,amssymb,amsfonts,twocolumn,nofootinbib,longbibliography]{revtex4}
\usepackage{}
\usepackage{amssymb}
\usepackage{bm,mathrsfs}
\usepackage{graphicx}
\usepackage{epsfig}
\usepackage{amsmath,bbm}
\usepackage{amsfonts,amssymb}
\usepackage{times}
\usepackage{verbatim}
\usepackage[sort&compress]{natbib}
\usepackage{amsmath}
\usepackage{bm}
\usepackage{float}
\usepackage{textgreek}
\usepackage{textcomp}
\allowdisplaybreaks[4]
\usepackage[colorlinks,breaklinks,linkcolor=blue,anchorcolor=blue,citecolor=blue,urlcolor=magenta,dvipdfm]{hyperref}

\usepackage{color}
\definecolor{zzz}{rgb}{0.9,0.0,0.4}

\begin{document}

\title{Non-Markovian Entanglement via Parametrically Mediated Molecular Cavity Optomechanics}
\author{Wenyao Hu,$^{1}$ H. N. Liu,$^{1}$ X. C. Zhang,$^{1}$ Cheng Shang,$^{2}$ Yan-Hui Zhou,$^{3,}$\footnote{\textcolor{zzz}{Corresponding author: yanhuizhou@126.com }}  and H. Z. Shen$^{1,}$\footnote{\textcolor{zzz}{Corresponding author: shenhz458@nenu.edu.cn }}}
\affiliation{$^1$Center for Quantum Sciences and School of Physics, Northeast Normal University, Changchun 130024, China\\
$^2$Analytical Quantum Complexity RIKEN Hakubi Research Team,
RIKEN Center for Quantum Computing (RQC), 2-1 Hirosawa, Wako, Saitama 351-0198, Japan\\
$^3$Quantum Information Research Center and Jiangxi Province Key Laboratory of Applied Optical Technology, Shangrao Normal University, Shangrao 334001, China}

\begin{abstract}
Quantum entanglement in molecular cavity optomechanical (COM) systems serves as a fundamental resource to explore quantum-enhanced sensing, coherent molecular dynamics, and quantum information processing, which has been studied under Markovian approximation but not yet explored in non-Markovian reservoirs. In this paper, we propose a scheme to manipulate quantum entanglement in molecular COM system coupling with degenerate optical parametric amplifier (DOPA) under Markovian approximation. We reveal that both strength and phase of DOPA can be tuned to enhance optical-vibration entanglement and suppress vibration-vibration entanglement. The optical-vibration entanglement can exist at room temperature, demonstrating strong robustness against thermal noise. Moreover, we extend results from Markovian to non-Markovian regimes, which can further enhance both types of entanglement with DOPA fixed. In particular, we find that non-Markovian effects can reverse influences of DOPA on vibration-vibration entanglement from suppression to enhancement, leading to regeneration of entanglement. The scheme provides an effective way to control and enhance quantum entanglement in molecular COM system through steering DOPA and non-Markovian effects.
\end{abstract}



\maketitle
\section{Introduction}
Quantum entanglement featuring inseparability of particles regardless of their distance \cite{RHorodecki818652009} is a key resource in quantum networks \cite{YRen1621005862022}, quantum communication \cite{FXu920250022020,RUrsin34812007}, and remote state preparation \cite{DHan1723001032023}, breaking through standard quantum limit in quantum sensing \cite{CLDegen890350022017,LPezze900350052018,ZZhang1141105062015,YXia174702023,JHuang110313022024} and computing \cite{TMGraham6044572022}. Entanglement has been demonstrated in numerous quantum systems, such as ions \cite{VKrutyanskiy1300508032023}, atomic ensembles \cite{XYLuo1290505032022,TvanLeent607692022}, photons \cite{XCYao62252012,ZGe10eadm75652024}, superconducting circuits \cite{PCampagneIbarcq1202005012018,PKurpiers5582642018,WNing1230605022019}, and various COM devices \cite{PChen70131612025,TAPalomaki3427102013,RRiedinger5303132016,SMancini881204012002,DVitali980304052007,LTian1102336022013,YDWang1102536012013,SBarzanjeh5704802019,YFJiao1251436052020,DGLai1290636022022,JHuang1060635062022,RRiedinger5564732018,CFOckeloenKorppi5564782018,ZYFan1722008662023}. By utilizing efficient light-motion couplings \cite{MAspelmeyer8613912014}, COM systems have contributed to many applications, such as quantum transducers \cite{FLecocq1160436012016,ZShen1292436012022}, quantum memory \cite{MBKristensen1321008022024}, ground-state cooling \cite{JDTeufel4753592011,DGLai1020115022020,YMLiu1130135232026}, weak-force sensing \cite{BBLi1027992021,EGavartin75092012,DMason157452019}, phonon lasing \cite{ISGrudinin1040839012010,GWang5732017,JZhang124792018,HJing1130536042014}, optomechanically induced transparency \cite{SWeis33015202010,AHSafaviNaeini472692011}, and squeezed or entangled states \cite{EEWollman3499522015,YFJiao23011542024}.

Compared to conventional optomechanics, molecular COM systems \cite{PRoelli111642016,REsteban5518892022,FBenz3547262016,PRoelli1144862024,MKSchmidt1062912016,MKDezfouli412452017,ALombardi80110162018,TNeuman1000434222019,YZhang716762020,SHughes1040454312021,
IShlesinger835062021,LAJakob1432912023,AKoner1477532023} offer three advantages: (i) larger optomechanical coupling facilitated by plasmonic nanocavities with subwavelength volumes, (ii) higher vibration frequency, and (iii) stronger collective effects due to characteristics of molecules. These specific advantages bring potential applications including heat transfer \cite{SMAshrafi1000138262019}, collective quantum entanglement \cite{JHuang1101843062024,HYYu1360136022026,EKBerinyuy1408062025,EKongkuiBerinyuy2051178202026}, molecular induced transparency \cite{BYin1110435072025}, robust photon blockade \cite{JTang2512210352025}, frequency up-conversion \cite{FZou1321536022024,WChen37412642021,AXomalis37412682021}, mechanical lasing \cite{MKSchmidt260330412024}, measurement \cite{JLiu68672018,JLiu54502017,MTabatabaei27522015}, molecule-protein analysis \cite{TSadhanasatish9eadg33472023}, nonlinear effect \cite{FBenz3547262016,ALombardi80110162018}, photonic spin Hall effect \cite{AkhterMunir1120235172025}, and phonon laser \cite{Yin711912026}.
Optomechanics has proposed schemes to enhance the performance of COM systems by parametric amplification  \cite{SPinaOtey930338352016,HXiong5812015,YWang170240092022,WZhang1090437122024}. DOPA is placed inside driven optomechanical cavity, which can lead to amplification and modulate optomechanical coupling  \cite{CSHu1000438242019,HHAdamyan920538182015,XYLu1140936022015,SSavelev850138112012}. It offers a mechanism to manipulate dynamical instabilities and nonlinear effects, bringing about enhanced control with applications, such as quantum squeezing and entanglement \cite{BHussain1050637042022,Wu260}, enhanced optomechanical cooling \cite{SHuang790138212009}, normal-mode splitting \cite{SHuang800338072009}, photon blockade \cite{Wu2025PRA}, and atom-cavity interactions \cite{WQin1200936012018}.

The Markovian approximation remains valid only when coupling between system and reservoir is weak, provided that characteristic time of bath is much smaller than that of system. When these are not met, Markovian approximation becomes restricted, necessitating us to consider influences of non-Markovian effects on system dynamics~\cite{SLonghi740638262006,AJLeggett5911987,CAddis900521032014,LXin1050537062022,SLorenzo880201022013,ARivas1050504032010,SLuo860441012012,
MMWolf1011504022008,XMLu820421032010,DChruscinski1121204042014,WZZhang930638532016,JFTriana1161836022016,HZShen980621062018,KSinha1240436032020,Song103L0106012021,WWu103L0106012022,Song103L0106012023,liu103L0106012022,HZShen930121072016,ZYLi1090237122024,
JLLi4756142022,HZShen990321012019,HZShen27315042019,HZShen4328522018,HZShen960338052017,HZShen970421212018}. Moreover, non-Markovian effects play meaningful roles in quantum optics including quantum control, quantum channels, and quantum state engineering~\cite{FCaruso8612032014,RLoFranco900543042014,SBXue860523042012}, which have been confirmed by experimental realization~\cite{BHLiu79312011,SGroblacher676062015,SJXiong1000321012019,SCialdi1000521042019,DKhurana990221072019,KHMadsen1062336012011,YGuo1262304012021,
BWLi1291405012022,UHoeppe1080436032012,JSXu820423282010,SAUriri1010521072020,MHAnderson4732021993,ZDLiu1020622082020,
HPBreuer880210022016,IdeVega890150012017}.
The above considerations lead us to investigate these issues: (i) Can quantum entanglement be controlled by DOPA in molecular COM systems? (ii) How can non-Markovian effects enhance collective entanglement? (iii) Can non-Markovian effects make entanglement robust against thermal noise?

In this paper, we present a molecular COM scheme incorporating DOPA under Markovian approximation, focusing on how DOPA affects bipartite entanglement. These results show that DOPA enhances optical-vibration entanglement and suppresses vibration-vibration entanglement. The optical-vibration entanglement maintains robustness to thermal noise and remains observable under room temperature conditions. Moreover, we extend study of entanglement to non-Markovian regimes and compare results with those under Markovian approximation. In non-Markovian reservoir, optical-vibration entanglement still exists and is further enhanced, regardless of presence of DOPA. We find that non-Markovian effects reverse influences of DOPA on vibration-vibration entanglement from suppression to enhancement (regeneration of entanglement).

The paper is organized as follows. In Sec.~\ref{two}, we construct a system model with DOPA and derive motion equations in Markovian reservoir. In Sec.~\ref{three}, we discuss influences of DOPA on entanglement. In Sec.~\ref{four}, we extend entanglement from Markovian to non-Markovian regimes. In Sec.~\ref{five}, we explore influences of non-Markovian effects on bipartite entanglement. The main conclusions are summarized in Sec.~\ref{six}.
\begin{figure}[t]
\centerline{
\includegraphics[width=7.2cm, height=3.0cm, clip]{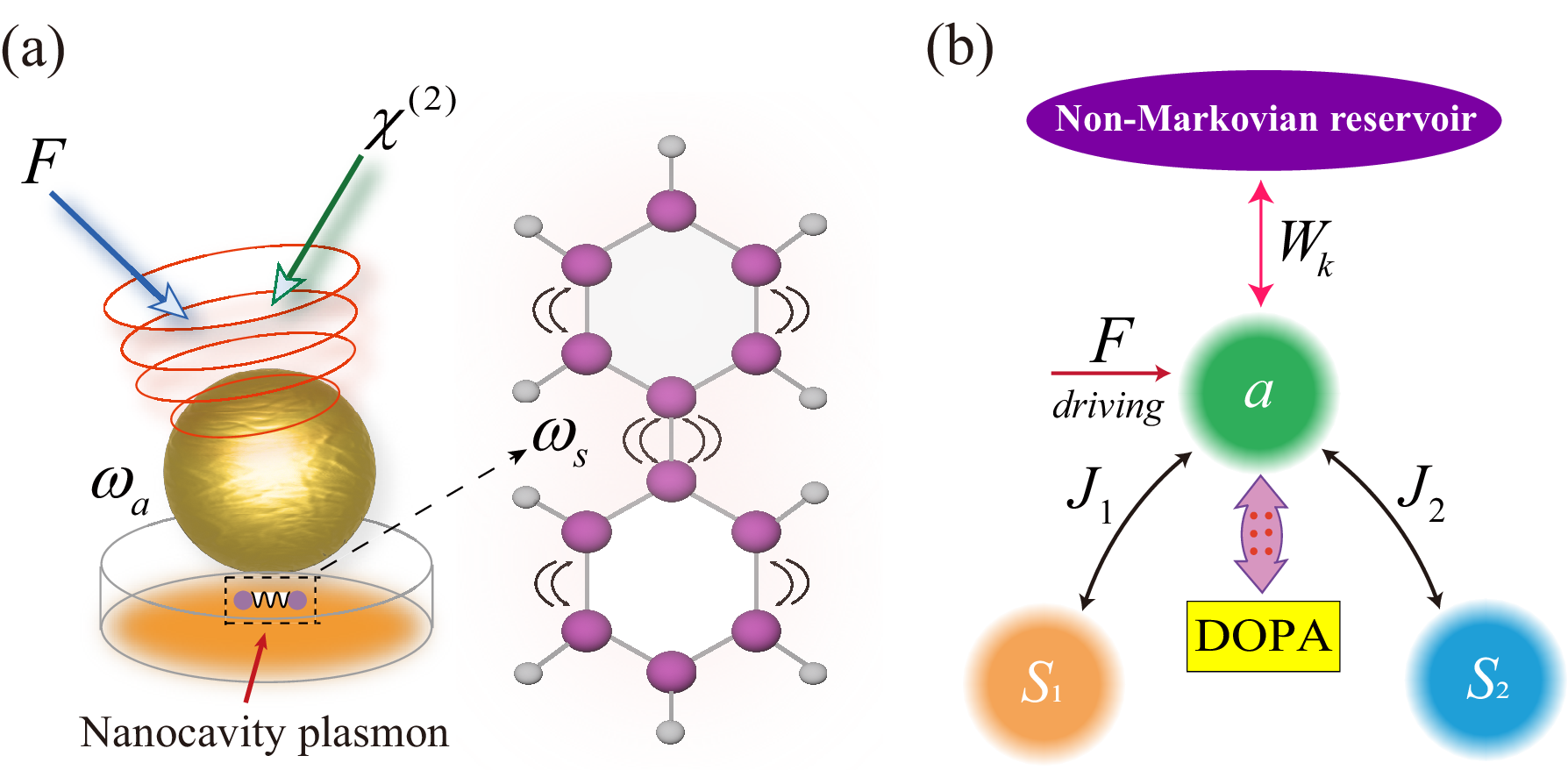}}
\caption{(a) Schematic diagram of molecular COM system model (left) and molecular vibration diagram (right). The metallic nanoparticle and molecules with vibrational frequency $\omega_s$ are deposited on a metallic substrate to form a plasmonic nanocavity with frequency $\omega_a$. The system is driven by external laser field with amplitude $F$ and incorporates DOPA produced by $\chi^{(2)}$ nonlinear medium. (b) Optomechanical couplings (${J_1}$, ${J_2}$) between cavity mode $a$ and collective molecular vibrational modes (${S_1}$, ${S_2}$). The cavity mode $a$ couples to one non-Markovian reservoir with coupling strength ${W_k}$. Markovian approximation cases are presented in Sec.~\ref{two} and Sec.~\ref{three}, whereas Sec.~\ref{four} and Sec.~\ref{five} cover non-Markovian reservoir.}\label{fig1ableftright}
\end{figure}
\section{MODEL AND DYNAMICS UNDER MARKOVIAN APPROXIMATION}\label{two}
We consider a molecular COM system consisting of a plasmonic nanocavity and $Q$ identical molecules incorporating DOPA (producing by $\chi^{(2)}$ nonlinear medium). Plasmonic nanocavities are essential to realize strong coupling in molecular COM systems due to their ability to confine light to nanoscale volumes \cite{PRoelli1144862024}, enhancing light-matter interactions. The plasmonic system can be realized using nanoparticle-on-mirror configuration \cite{SMubeen1220882012}, facilitated by scanning-tunneling-microscope tip positioned above metallic substrate \cite{RZhang498822013}, where $Q$ molecules are placed near metallic nanostructures acting as nanoresonators in Fig.~\ref{fig1ableftright}(a). The vibrational mode of each molecule is modeled as a harmonic oscillator characterized by resonance frequency ${\omega _s}$ in the limit of low excitation. The corresponding normal coordinate operator is defined as $\hat{r} = r_{0}(\hat{s}^\dagger + \hat{s})$, where $r_{0}$ represents zero-point fluctuation amplitude, while $\hat{s}$ ($\hat{s}^{\dagger}$) denotes annihilation (creation) operator. We assume that cavity resonance frequency $\omega_a$ is off-resonant with allowed electronic transitions of molecules. Consequently, the interaction between cavity field and molecular vibrations is parametric, where mechanical motion modulates optical cavity resonance frequency with $\omega_a(r) \approx \omega_a - \tilde{J}_s r$. The coupling strength is given by $\tilde{J}_s = \omega_a [\partial \Xi (r) / \partial r] / (\epsilon_0 \mathcal{V}_a)$, where $\Xi (r)$, $\epsilon_0$, and $\mathcal{V}_a$ are molecular polarizability, vacuum permittivity, and volume of cavity \cite{PRoelli111642016}. We first study Markovian case and subsequently discuss influences of non-Markovian effects on entanglement.
The total Hamiltonian in the rotating frame of system reads (setting $\hbar  \equiv 1$)
\begin{equation}
\begin{aligned}
\hat{H}_{M}=&\Delta_{a}\hat{a}^{\dagger}\hat{a}+\sum_{l=1}^{2}\omega_s\hat{S}_{l}^{\dagger}\hat{S}_{l}+\sum_{l=1}^{2}J_{l}\hat{a}^{\dagger}\hat{a}(\hat{S}_{l}^{\dagger}+\hat{S}_{l})\\
&+iF(\hat{a}^{\dagger}-\hat{a})+i{\cal G}(e^{i\theta}\hat{a}^{\dagger 2}-e^{-i\theta}\hat{a}^{2}),
\label{HamiltonianEnd}
\end{aligned}
\end{equation}
where $\hat{S}_{1}=\sum_{n=1}^{\Lambda}\hat{s}_{n}/\sqrt{\Lambda}$ and $\hat{S}_{2}=\sum_{n=\Lambda+1}^{Q}\hat{s}_{n}/\sqrt{Q-\Lambda}$ with $[\hat{S}_l, \hat{S}_{l'}^\dagger] = \delta_{ll'}$ ($l = 1,2$) denote $Q$ molecules collective operators, where $\Lambda$ is the number of molecules included in ${S_1}$ collective mode. $\hat{a}$ ($\hat{a}^{\dagger}$) and $\hat{s}_{n}$ ($\hat{s}_{n}^{\dagger}$) are respectively annihilation (creation) operators of cavity mode and vibrational mode of $n$th molecule. ${J_1} = {J_s}\sqrt \Lambda $ and ${J_2} = {J_s}\sqrt {Q - \Lambda}$ are collective optomechanical coupling strengths. $J_s = -\tilde{J}_s r_{0}$ represents molecular optomechanical coupling constant. $F$ in Eq.~(\ref{HamiltonianEnd}) is amplitude of driving field with frequency ${\omega _L}$. The last term in Eq.~(\ref{HamiltonianEnd}) describes coupling of intracavity field with DOPA. ${\cal G}$ and $\theta$ respectively denote strength and phase of DOPA with frequency ${\omega _p}$, where ${\omega _p} = 2{\omega _L}$ has been used. ${\Delta _a} = {\omega _a} - {\omega _L}$ represents detuning between cavity and driving field. The optomechanical coupling strength is derived as $J_s = -\omega_a I_s \sqrt{\hbar / (2 m \omega_s)} / (\epsilon_0 \mathcal{V}_a)$, where $I_s = \partial \Xi (r) / \partial r$ and $\Xi (r)$ is Raman polarizability. $m$ denotes the effective mass  of vibrations. To provide a quantitative estimation based on Refs.~\cite{PRoelli111642016,MKSchmidt1062912016,REsteban5518892022,FBenz3547262016,ALombardi80110162018,RZhang498822013}, we adopt parameters $\omega_a/2\pi = 333\ \text{THz}$, $\omega_s/2\pi = 29.9\ \text{THz}$, $I_s^2/m \approx 3 \times 10^{-10} \epsilon_0^2 \, \text{\AA}^4 \, \text{amu}^{-1}$, $\mathcal{V}_a = 1.5 \times 10^{-6} \ $\textmu$\text{m}^3$, $-J_s/2\pi \sim 10\text{--}100 \ \text{GHz}$, and $\omega_s/2\pi \sim 6\text{--}48 \ \text{THz}$. Apart from free Hamiltonians of two collective vibrational modes, Eq.~(\ref{HamiltonianEnd}) includes terms of $Q - 2$ dark modes. However, these $Q - 2$ dark modes are decoupled from both cavity and collective vibrational modes. Consequently, they do not participate in dynamics of system and remain in their respective thermal states \cite{DGLai1020115022020}. We neglect these terms associated with non-interacting dark modes from $\hat{H}_{M}$ in Eq.~(\ref{HamiltonianEnd}).

We linearize the Heisenberg-Langevin equations with Eq.~(\ref{HamiltonianEnd}) by expressing each operator as a sum of steady-state values and quantum fluctuation operators $\hat{a} = \alpha_a + \delta\hat{a}$, $\hat{S}_1 = \beta_1 + \delta\hat{S}_1$, and $\hat{S}_2 = \beta_2 + \delta\hat{S}_2$ in Eqs.~(\ref{Langevinequations1})-(\ref{Langevinequations3}). The steady-state solutions in Eqs.~(\ref{mean1})-(\ref{mean3}) representing mean-field amplitudes of cavity and vibrational modes are ${\alpha _a} = F/[(i\Delta  + \kappa ) - 2{\cal G} {e^{i\theta }}]$, ${\beta _1} = { - i{J_1}{{\left| {{\alpha _a}} \right|}^2}}/{(i{\omega _s} + {\gamma _1})}$, and ${\beta _2} = { - i{J_2}{{\left| {{\alpha _a}} \right|}^2}}/{(i{\omega _s} + {\gamma _2})}$, where $\Delta = \Delta_a + \sum_{l=1}^{2} J_l (\beta_l^* + \beta_l)$. Although steady-state amplitude $\alpha_a$ is complex, we can choose an appropriate phase reference for cavity field such that $\alpha_a$ becomes real. The fluctuation operators satisfy Eqs.~(\ref{linearizedLangevinequations1})-(\ref{linearizedLangevinequations3}).
Defining quadrature fluctuation operators vector $\hat{U}^M(t) = (\delta \hat{X}_{a}, \delta \hat{Y}_{a}, \delta \hat{O}_{1}, \delta \hat{P}_{1}, \delta \hat{O}_{2}, \delta \hat{P}_{2})^{\mathrm{T}}$ and noise operators vector $\hat{B}^M(t) = (\sqrt{2\kappa} \hat{X}_a^{{in}}, \allowbreak \sqrt{2\kappa} \hat{Y}_a^{{in}}, \allowbreak \sqrt{2\gamma_1} \hat{O}_1^{{in}}, \allowbreak \sqrt{2\gamma_1} \hat{P}_1^{{in}}, \allowbreak \sqrt{2\gamma_2} \hat{O}_2^{{in}}, \allowbreak \sqrt{2\gamma_2} \hat{P}_2^{{in}})^{\mathrm{T}}$ with $\delta \hat{X}_{a}=(\delta\hat{a}+\delta\hat{a}^{\dagger})/\sqrt{2}, \delta \hat{Y}_{a}=i(\delta\hat{a}^{\dagger}-\delta\hat{a})/\sqrt{2}$, $\delta \hat{O}_{1}=(\delta\hat{S}_{1}+\delta\hat{S}_{1}^{\dagger})/\sqrt{2}, \delta \hat{P}_{1}=i(\delta\hat{S}_{1}^{\dagger}-\delta\hat{S}_{1})/\sqrt{2}$, $\delta \hat{O}_{2}=(\delta\hat{S}_{2}+\delta\hat{S}_{2}^{\dagger})/\sqrt{2}, \delta \hat{P}_{2}=i(\delta\hat{S}_{2}^{\dagger}-\delta\hat{S}_{2})/\sqrt{2}$, $\hat{X}_{a}^{in}=(\hat{a}^{in}+\hat{a}^{in\dagger})/\sqrt{2}, \hat{Y}_{a}^{in}=i(\hat{a}^{in\dagger}-\hat{a}^{in})/\sqrt{2}$, $\hat{O}_{1}^{in}=(\hat{S}_{1}^{in}+\hat{S}_{1}^{in\dagger})/\sqrt{2}, \hat{P}_{1}^{in}=i(\hat{S}_{1}^{in\dagger}-\hat{S}_{1}^{in})/\sqrt{2}$, $\hat{O}_{2}^{in}=(\hat{S}_{2}^{in}+\hat{S}_{2}^{in\dagger})/\sqrt{2}$, and $\hat{P}_{2}^{in}=i(\hat{S}_{2}^{in\dagger}-\hat{S}_{2}^{in})/\sqrt{2}$ (see Appendix \ref{Der1} for  operators definition and details), Eqs.~(\ref{linearizedLangevinequations1})-(\ref{linearizedLangevinequations3}) can be written as $\dot{\hat{U}}^M(t) = A^M \hat{U}^M(t) + \hat{B}^M(t)$ with drift matrix
\begin{equation}
\resizebox{0.9\hsize}{!}{$
A^M = \left( {\begin{array}{*{20}{c}}
{2{\cal G} \cos \theta  - \kappa }&{2{\cal G} \sin \theta  + \Delta }&0&0&0&0\\
{2{\cal G} \sin \theta  - \Delta }&{ - 2{\cal G} \cos \theta  - \kappa }&{ - 2{G_1}}&0&{ - 2{G_2}}&0\\
0&0&{ - {\gamma _1}}&{{\omega _s}}&0&0\\
{ - 2{G_1}}&0&{ - {\omega _s}}&{ - {\gamma _1}}&0&0\\
0&0&0&0&{ - {\gamma _2}}&{{\omega _s}}\\
{ - 2{G_2}}&0&0&0&{ - {\omega _s}}&{ - {\gamma _2}}
\end{array}} \right)
$}.
\label{driftmatrix}
\end{equation}
According to Routh-Hurwitz criterion, system is stable and reaches steady state only when all eigenvalues of $A^M$ in Eq.~(\ref{driftmatrix}) have negative real parts. Due to Gaussian properties of quantum noise and linearity of quantum Langevin equations, quantum state of system is characterized by $6 \times 6$ covariance matrix $V^M$ with matrix elements $V^M_{ij} = \langle U_{i}^{M}(\infty)U_{j}^{M}(\infty) + U_{j}^{M}(\infty)U_{i}^{M}(\infty)\rangle / 2$ ($i, j = 1, 2, \cdots, 6$). The steady-state covariance matrix $V^M$ is determined by Lyapunov equation $A^M V^M + V^M (A^M)^{\mathrm{T}} = -D^M$, where $D^M = \text{diag}[\kappa, \kappa, (2\bar{N}_1 + 1)\gamma_1, (2\bar{N}_1 + 1)\gamma_1, (2\bar{N}_2 + 1)\gamma_2, (2\bar{N}_2 + 1)\gamma_2]$ represents diffusion matrix with matrix elements $D_{ij}^M$ defined by $\langle \hat{B}_i^M(t)\hat{B}_j^M(t') + \hat{B}_j^M(t')\hat{B}_i^M(t) \rangle/2 = D_{ij}^M \delta(t-t')$.
\begin{table*}[t]
\centering
\caption{Comparison of experimentally feasible parameters between conventional COM and current molecular COM systems.}
\label{tab:comparison}
\renewcommand{\arraystretch}{1.4} 
\begin{tabular*}{\textwidth}{@{\extracolsep{\fill}}lccc}
\hline \hline
Quantity & Symbol & Conventional COM \cite{SWeis33015202010} & Molecular COM \cite{PRoelli111642016,WChen37412642021,FZou1321536022024,JHuang1101843062024,AXomalis37412682021,AkhterMunir1120235172025,BYin1110435072025} \\
\hline
Cavity resonance frequency                    & $\omega_a$ & 193 THz                 & 460 THz                        \\
Molecular optomechanical coupling strength    & $J_s$          & 465.3 Hz                & $2\pi \times 30$ GHz           \\
Molecular vibrational frequency               & $\omega_s$     & $2\pi \times 23$ MHz    & $2\pi \times 30$ THz           \\
Cavity dissipation rate                       & $\kappa$       & $\sim 6.43$ MHz         & $\sim 2\pi \times 10$ THz           \\
Collective vibrational modes dissipation rate & $\gamma_{1,2}$ & 0.24 MHz                & $2\pi \times 0.03$ THz         \\
Temperature                                   & $T$            & 10 mK                   & 312 K                          \\
Laser driving frequency                          & $\omega_L$     & 193 THz                 & 300 THz                        \\
\hline \hline
\end{tabular*} 
\end{table*}
To quantify bipartite entanglement between cavity and collective vibrational modes, Markovian logarithmic negativity ${E^M}$ \cite{GVidal650323142002,MBPlenio950905032005,RSimon8427262000} is defined as entanglement measure
\begin{equation}
\begin{aligned}
E^M = \max[0, -\ln(2\xi^M)],
\label{DSFX}
\end{aligned}
\end{equation}
where $\xi^M = 2^{-1/2} \{ \mathcal{Q} - [ \mathcal{Q}^2 - 4 \det({V^M_{sub}}) ]^{1/2} \}^{1/2}$ is minimum symplectic eigenvalue of matrix ${V^M_{sub}}$ with $\mathcal{Q} = \det(\phi_1) + \det(\phi_2) - 2 \det(\phi_3)$. $\phi_1, \phi_2$, and $\phi_3$ are $2 \times 2$ submatrices derived from covariance matrix ${V^M_{sub}}$. For a subsystem composed of modes 1 and 2, covariance matrix can be written as ${V^M_{sub}} = \left( \begin{smallmatrix} \phi_1 & \phi_3 \\ \phi_3^\mathrm{T} & \phi_2 \end{smallmatrix} \right)$.
We focus on quantifying bipartite entanglement between cavity $a$ and vibrational modes ${S_{1,2}}$ ($E_{a{S_l}}^M$) as well as between two vibrational modes ${S_1}$ and ${S_2}$ ($E_{S_1{S_2}}^M$).
It is important to clarify how Eq.~\eqref{DSFX} is applied to calculate entanglement. Although mathematical definition $E^M$ in Eq.~\eqref{DSFX} remains the same, the difference stems from selection of $4\times 4$ subsystem covariance matrix $V_{sub}^M$ from whole system covariance matrix $V^M$. Based on definition of fluctuation operator vector $U^M(t)$, (i) we quantify $E_{aS_l}^M$ by extracting rows and columns of matrix elements corresponding to cavity mode $a$ ($\delta \hat{X}_{a}$, $\delta \hat{Y}_{a}$) and collective vibrational mode $S_l$ ($\delta \hat{O}_{1},\delta \hat{P}_{1}$ or $\delta \hat{O}_{2},\delta \hat{P}_{2}$). (ii) In parallel, we obtain submatrix for $E_{S_1S_2}^M$ by using elements with $S_1$ ($\delta \hat{O}_{1},\delta \hat{P}_{1}$) and $S_2$ ($\delta \hat{O}_{2},\delta \hat{P}_{2}$). We only present results for $E_{a{S_2}}^M$ and $E_{S_1{S_2}}^M$ as representative cases because $E_{a{S_1}}^M$ exhibits similar properties.
\section{Discussions on DOPA mediated bipartite Entanglement UNDER MARKOVIAN APPROXIMATION}\label{three}
In this section, we present numerical analysis results under Markovian approximation, exploring bipartite entanglement properties of proposed molecular COM system. We choose experimentally feasible
parameters, i.e., \(\omega_L/2\pi = 300\ \text{THz}\), \(\omega_s/2\pi = 30\ \text{THz}\), \(\kappa/\omega_s = 0.3\), $\gamma_1/\omega_s = \gamma_2/\omega_s = 10^{-4}$, \(J_s/2\pi = 30\ \text{GHz}\), \(\Lambda = 0\), \(Q = 100\), \(\theta = \pi /2\), and \(T \approx 312\ \text{K}\) \cite{JHuang1101843062024,JLiu54502017,FZou1321536022024,MKSchmidt260330412024,AkhterMunir1120235172025,EKBerinyuy1408062025}. Table~\ref{tab:comparison} compares system parameters to conventional COM and current molecular COM, where molecular COM systems possess extremely large optomechanical coupling $J_s/2\pi = 30$ GHz (corresponding to 465.3 Hz in conventional COM) and ultrahigh vibrational frequency $\omega_s/2\pi = 30$ THz (corresponding to $2\pi \times 23$ MHz in conventional COM), resulting in very low thermal phonon number $\bar{N}_{th} = 0.01$ at $T \approx 312$ K.
\begin{figure}[h]
   \centerline{
   \includegraphics[width=7.8cm, height=5.55cm, clip]{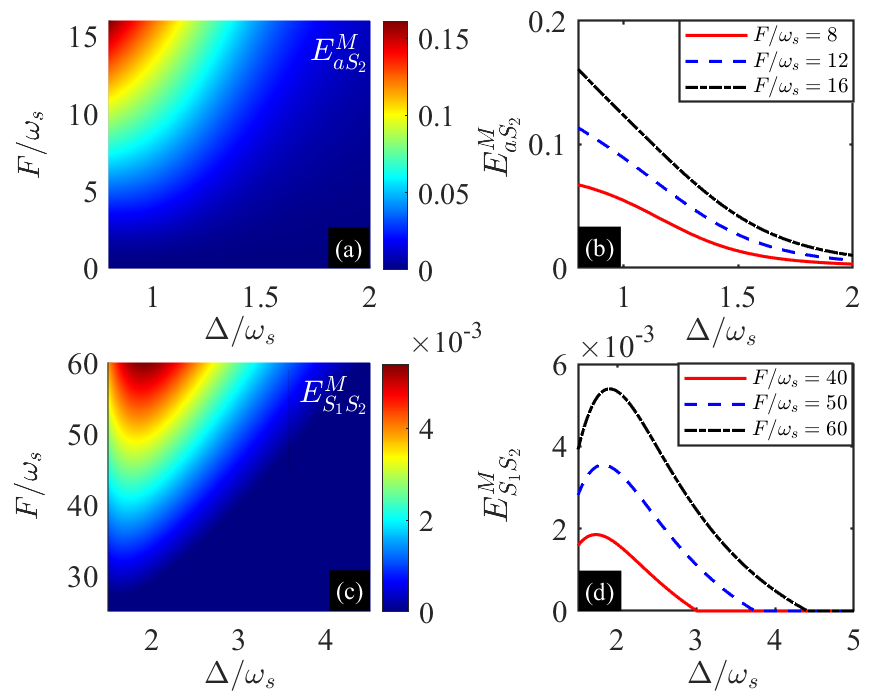}}
   \caption{Markovian logarithmic negativity in Eq.~\eqref{DSFX} for (a,b) optical-vibration entanglement $E_{aS_2}^{M}$ and (c,d) vibration-vibration entanglement $E_{S_1S_2}^{M}$ is plotted as a function of detuning $\Delta$ and driving amplitude $F$. The red solid, blue dashed, and black dash-dotted lines respectively correspond to $F /{\omega _s} = 8, 12,$ and $16$ in (b), while $F /{\omega _s} = 40, 50,$ and $60$ in (d). The parameters chosen are $\omega_s/2\pi = 30\,\text{THz}$, $J_s/2\pi = 30\,\text{GHz}$, $\kappa/\omega_s = 0.3$, ${\cal G}/\omega_s = 0$, $Q = 100$, $\theta = \pi/2$ with $\gamma_1/\omega_s = \gamma_2/\omega_s = 10^{-4}$, $\Lambda = 0$, $\bar{N}_{th} = 0.01$ for (a,b) and $\gamma_1/\omega_s = \gamma_2/\omega_s = 0.3$, $\Lambda = 50$, $\bar{N}_{th} = 0.001$ for (c,d).}\label{fig2abcd}
 \end{figure}

In Fig.~\ref{fig2abcd}, we plot bipartite entanglement as a function of detuning $\Delta$ for different driving amplitudes $F$ without participation of DOPA (${\cal G}= 0$).
In Fig.~\ref{fig2abcd}(a), optical-vibration entanglement $E_{aS_2}^M$ increases with driving amplitude $F$ and reaches peak value 0.1611 around detuning $\Delta/\omega_s \approx 0.8$ at maximum $F /{\omega _s} = 16$. Figure~\ref{fig2abcd}(b) shows $\Delta$ dependence with $F$ fixed. The red solid, blue dashed, and black dash-dotted lines respectively correspond to $F /{\omega _s}$ = 8, 12, and 16. We find that while $E_{aS_2}^M$ grows with increasing $F$, it decreases with increasing $\Delta$.
In Fig.~\ref{fig2abcd}(c), vibration-vibration entanglement $E_{S_1S_2}^M$ grows as $F$ increases and reaches peak value 0.0054 around $\Delta/\omega_s \approx 2$ at maximum $F /{\omega _s} = 60$. Specifically, vibrational modes are uncorrelated ($E_{S_1S_2}^M\approx0$) below a driving threshold of about $F/\omega_s = 30$, indicating that strong optomechanical coupling is required to realize entanglement between two vibrational modes. Figure~\ref{fig2abcd}(d) plots $E_{S_1S_2}^M$ as a function of detuning $\Delta$ for three driving amplitudes $F$. The red solid, blue dashed, and black dash-dotted lines respectively correspond to $F /{\omega _s}$ = 40, 50, and 60. In all cases, $E_{S_1S_2}^M$ first rises to maximum and then decreases as $\Delta$ increases. These results show that both types of entanglement can be tuned by $\Delta$ and $F$. The optical-vibration entanglement $E_{aS_2}^M$ is accessible at lower $\Delta$ and $F$, whereas vibration-vibration entanglement $E_{S_1S_2}^M$ requires much higher ones.
 \begin{figure}[thpb]
   \centerline{
   \includegraphics[width=7.8cm, height=5.6cm, clip]{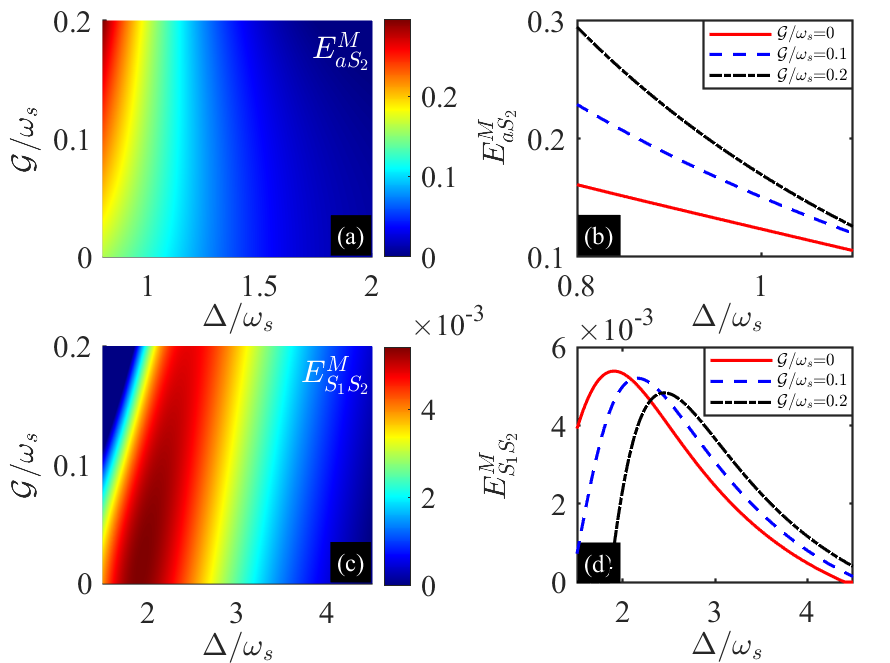}}
   \caption{Markovian logarithmic negativity in Eq.~\eqref{DSFX} for (a,b) optical-vibration entanglement $E_{aS_2}^{M}$ and (c,d) vibration-vibration entanglement $E_{S_1S_2}^{M}$ is plotted as a function of detuning $\Delta$ and DOPA strength $\mathcal{G}$. The red solid, blue dashed, and black dash-dotted lines respectively correspond to $\mathcal{G} /\omega_s = 0, 0.1,$ and $0.2$. The parameters chosen are $F/\omega_s = 16$ for (a,b) and $F/\omega_s = 60$ for (c,d), while other parameters are the same as in Fig.~\ref{fig2abcd}.}\label{fig3abcd}
 \end{figure}

In Fig.~\ref{fig3abcd}, we plot optical-vibration entanglement $E_{aS_2}^M$ and vibration-vibration entanglement $E_{S_1S_2}^M$ as functions of cavity detuning $\Delta$ and DOPA strength ${\cal G}$ to explore influences of DOPA on entanglement. The red solid, blue dashed, and black dash-dotted lines respectively correspond to ${\cal G} /{\omega _s}$ = 0, 0.1, and 0.2.
In Fig.~\ref{fig3abcd}(a) and (b), \(E_{aS_2}^M\) exhibits significant enhancement ($E_{aS_2}^M=0.2949$) with increasing \({\cal G}\).
In Fig.~\ref{fig3abcd}(c) and (d), increasing $\mathcal{G}$ suppresses \(E_{S_1S_2}^M\). Specifically, the peak of $E_{S_1S_2}^M$ shifts to right as ${\cal G}$ increases from ${\cal G}/\omega_s = 0$ (red solid line) to ${\cal G}/\omega_s = 0.2$ (black dash-dotted line). These results show that while DOPA enhances optical-vibration entanglement, it suppresses vibration-vibration entanglement under Markovian approximation.
\begin{figure}[thpb]
   \centerline{
   \includegraphics[width=7.8cm, height=5.55cm, clip]{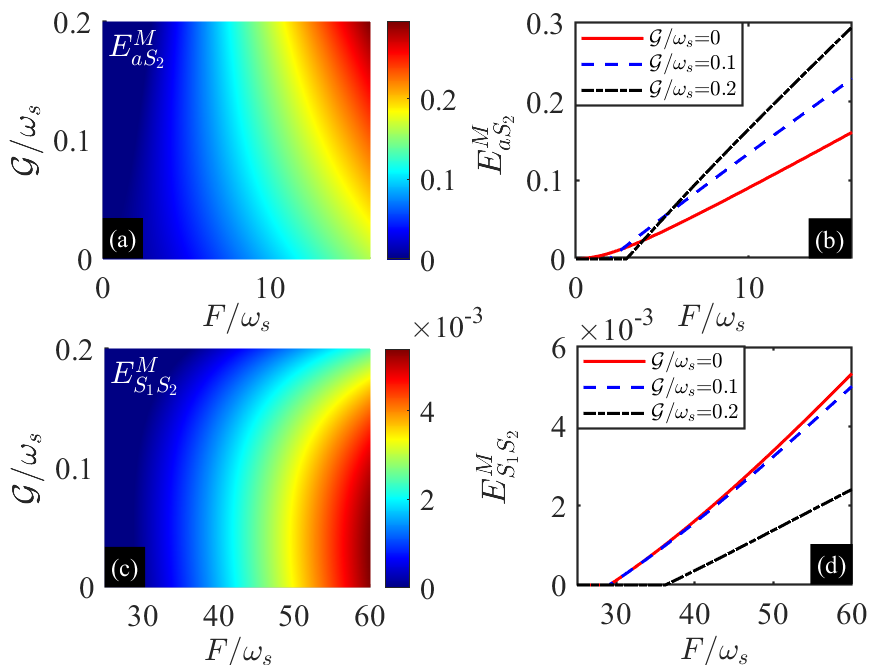}}
   \caption{Markovian logarithmic negativity in Eq.~\eqref{DSFX} for (a,b) optical-vibration entanglement $E_{aS_2}^{M}$ and (c,d) vibration-vibration entanglement $E_{S_1S_2}^{M}$ is plotted as a function of driving amplitude $F$ and DOPA strength $\mathcal{G}$. The red solid, blue dashed, and black dash-dotted lines respectively correspond to $\mathcal{G} /\omega_s = 0, 0.1,$ and $0.2$. The parameters chosen are $\Delta/\omega_s = 0.8$ for (a,b) and $\Delta/\omega_s = 2$ for (c,d), while other parameters are the same as in Fig.~\ref{fig2abcd}.}\label{fig4abcd}
 \end{figure}

We further consider influences of driving amplitude $F$ and DOPA strength ${\cal G}$ on bipartite entanglement in Fig.~\ref{fig4abcd}. The red solid, blue dashed, and black dash-dotted lines respectively correspond to ${\cal G} /{\omega _s}$ = 0, 0.1, and 0.2.
In Fig.~\ref{fig4abcd}(a) and (b), optical-vibration entanglement $E_{aS_2}^M$ exhibits obvious enhancement as $F$ and ${\cal G}$ increase. We find that $E_{aS_2}^M$ reaches maximum with strong driving $F=16$ and large DOPA strength ${\cal G}=0.2$. For a given $F$, $E_{aS_2}^M$ can be enhanced by increasing ${\cal G}$, confirming that DOPA increases entanglement between optical and vibrational modes.
In Fig.~\ref{fig4abcd}(c) and (d), while $E_{S_1S_2}^M$ grows with $F$, it is suppressed as ${\cal G}$ increases. This suppression becomes evident at higher driving powers.
\begin{figure}[thpb]
   \centerline{
   \includegraphics[width=7.8cm, height=5.55cm, clip]{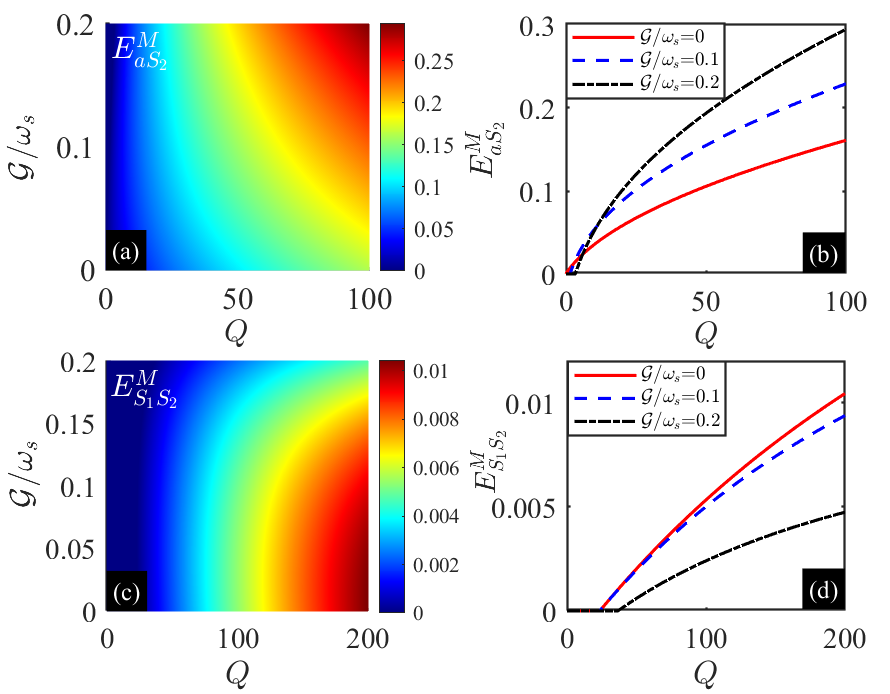}}
\caption{Markovian logarithmic negativity in Eq.~\eqref{DSFX} for (a,b) optical-vibration entanglement $E_{aS_2}^{M}$ and (c,d) vibration-vibration entanglement $E_{S_1S_2}^{M}$ is plotted as a function of molecular total number $Q$ and DOPA strength $\mathcal{G}$. The red solid, blue dashed, and black dash-dotted lines respectively correspond to $\mathcal{G} /\omega_s = 0, 0.1,$ and $0.2$. The parameters chosen are $F/\omega_s = 16$ with $\Delta/\omega_s = 0.8$ for (a,b) and $F/\omega_s = 60$, $\Lambda = Q/2$ with $\Delta/\omega_s = 2$ for (c,d), while other parameters are the same as in Fig.~\ref{fig2abcd}.}\label{fig5abcd}
 \end{figure}

\begin{figure}[thpb]
   \centerline{
   \includegraphics[width=7.8cm, height=5.55cm, clip]{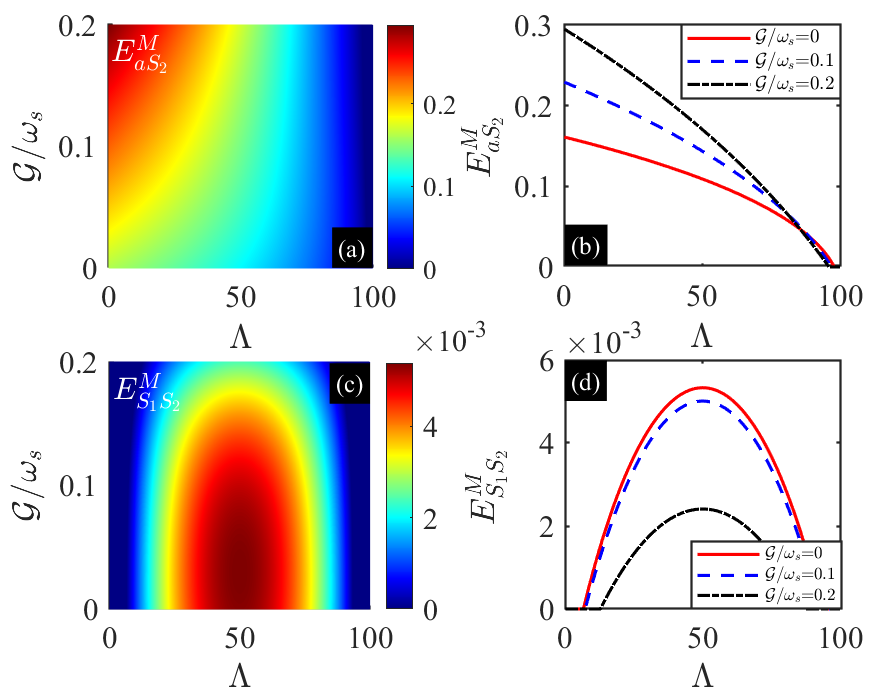}}
   \caption{Markovian logarithmic negativity in Eq.~\eqref{DSFX} for (a,b) optical-vibration entanglement $E_{aS_2}^{M}$ and (c,d) vibration-vibration entanglement $E_{S_1S_2}^{M}$ is plotted as a function of molecular collective mode distribution $\Lambda$ and DOPA strength $\mathcal{G}$. The red solid, blue dashed, and black dash-dotted lines respectively correspond to $\mathcal{G} /\omega_s = 0, 0.1,$ and $0.2$. The parameters chosen are $F/\omega_s = 16$ with $\Delta/\omega_s = 0.8$ for (a,b) and $F/\omega_s = 60$ with $\Delta/\omega_s = 2$ for (c,d), while other parameters are the same as in Fig.~\ref{fig2abcd}.}\label{fig6abcd}
 \end{figure}

To understand how collective effects of molecular ensembles affect bipartite entanglement with DOPA fixed, we examine influences of $E^{M}_{aS_2}$ and $E^{M}_{S_1S_2}$ on the total number of molecules \(Q\) and distribution \(\Lambda\) of molecules between two collective vibrational modes in Figs.~{\ref{fig5abcd}} and~{\ref{fig6abcd}}. The red solid, blue dashed, and black dash-dotted lines correspond to $\mathcal{G} /\omega _s$ = 0, 0.1, and 0.2, respectively.
In Fig.~\ref{fig5abcd}(a) and (b), $E^{M}_{aS_2}$ can be enhanced as the total number of molecules $Q$ increases. This is because collective coupling strength $G_2 = \sqrt{Q} G_s$ grows as $\sqrt{Q}$ at $\Lambda = 0$, which enhances entanglement between cavity mode $a$ and vibrational mode $S_2$. This enhancement originates from collective interactions of molecular ensembles, which leads to stronger collective effects.
In Fig.~\ref{fig5abcd}(c) and (d), $E^{M}_{S_1S_2}$ exhibits growth from 0 to larger value 0.0104 with increasing $Q$. The increase of total number $Q$ can enhance indirect coupling between two collective modes, transforming them from independence to entanglement.
We further investigate how the distribution number $\Lambda$ of molecules between two collective vibrational modes affects entanglement of system.
In Fig.~\ref{fig6abcd}(a) and (b), $E_{aS_2}^M$ gradually decreases as $\Lambda$ increases, which can be attributed to definition of effective collective coupling strength $G_2 = \sqrt{Q-\Lambda}G_s$ with $G_s = J_s|\alpha_a|$. All molecules contribute to mode $S_2$ at $\Lambda=0$, which leads to the strongest coupling $G_2 = \sqrt{Q}G_s$ and maximizes $E_{aS_2}^M$. As $\Lambda$ increases toward $Q$, the number of molecules contributing to mode $S_2$ decreases, eventually causing $G_2=0$ and $E_{aS_2}^M\approx 0$ at $\Lambda=Q=100$.
In contrast, Fig.~\ref{fig6abcd}(c) and (d) show that vibration-vibration entanglement $E_{S_1S_2}^M$ first rises and then decreases as $\Lambda$ increases from $0$ to $100$. Specifically, $E_{S_1S_2}^M$ reaches maximum when molecules are distributed between two collective vibrational modes at $\Lambda = Q/2=50$. This phenomenon can be attributed to the fact that entanglement between vibration modes $S_1$ and $S_2$ is induced by cavity field. Therefore, realizing this entanglement requires simultaneous strong coupling of cavity to both vibrational modes, which can be understood by analyzing interactions between two effective coupling strengths defined as $G_1 = \sqrt{\Lambda}G_s$ and $G_2 = \sqrt{Q-\Lambda}G_s$. The cavity decouples from first mode $S_1$ with $G_1=0$ at $\Lambda=0$, suppressing entanglement to mode $S_1$ and leading to negligible vibration-vibration entanglement ($E_{S_1S_2}^M \approx 0$). Similarly, the coupling to second mode $S_2$ becomes $G_2=0$ at $\Lambda=Q=100$, which decouples mode $S_2$ and suppresses vibration-vibration entanglement ($E_{S_1S_2}^M \approx 0$). Importantly, $E_{S_1S_2}^M$ reaches maximum when the number of molecules in two partitions is equal ($\Lambda=Q/2=50$), leading to balanced couplings $G_1=G_2$ that enhance entanglement between vibrational modes via cavity field.
\begin{figure}[thpb]
   \centerline{
   \includegraphics[width=7.8cm, height=5.55cm, clip]{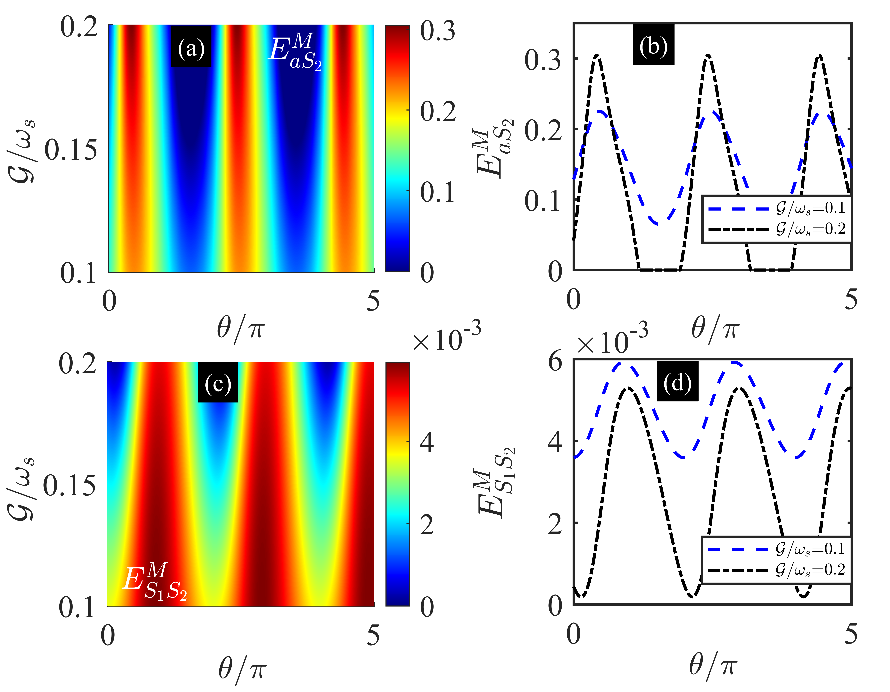}}
  \caption{Markovian logarithmic negativity in Eq.~\eqref{DSFX} for (a,b) optical-vibration entanglement $E_{aS_2}^{M}$ and (c,d) vibration-vibration entanglement $E_{S_1S_2}^{M}$ is plotted as a function of DOPA phase $\theta$ and strength $\mathcal{G}$. The blue dashed and black dash-dotted lines respectively correspond to ${\cal G}/\omega_s = 0.1$ and $0.2$. The parameters chosen are $F/\omega_s = 16$ with $\Delta/\omega_s = 0.81$ for (a,b) and $F/\omega_s = 60$ with $\Delta/\omega_s = 2$ for (c,d), while other parameters are the same as in Fig.~\ref{fig2abcd}.}\label{fig7abcd}
 \end{figure}

The phase $\theta$ of DOPA serves as another adjustable control parameter in our system. Figure~\ref{fig7abcd} illustrates bipartite entanglement as a function of DOPA phase $\theta$ and strength ${\cal G}$. The blue dashed and black dash-dotted lines respectively correspond to ${\cal G}/\omega_s = 0.1$ and $0.2$.
In Fig.~\ref{fig7abcd}(a) and (b), optical-vibration entanglement $E_{aS_2}^M$ exhibits periodic dependence on phase $\theta$. Specifically, $E_{aS_2}^M$ reaches peak values around $\theta \approx (2q + 1/2)\pi$ ($q$ is an integer), such as $0.5\pi$, $2.5\pi$, and $4.5\pi$. Conversely, $E_{aS_2}^M$ reaches minima around $\theta \approx (2q + 3/2)\pi$ (i.e., $1.5\pi$ and $3.5\pi$).
In Fig.~\ref{fig7abcd}(c) and (d), vibration-vibration entanglement $E_{S_1S_2}^M$ reaches peak values $E_{S_1S_2}^M = 0.0059$ around odd multiples of $\pi$, i.e., $\theta \approx (2q + 1)\pi$ (e.g., $\pi$, $3\pi$, and $5\pi$) and decreases to minima around even multiples of $\pi$, i.e., $\theta \approx 2q\pi$ (e.g., $0$, $2\pi$, and $4\pi$).

\begin{figure}[thpb]
   \centerline{
   \includegraphics[width=7.8cm, height=5.55cm, clip]{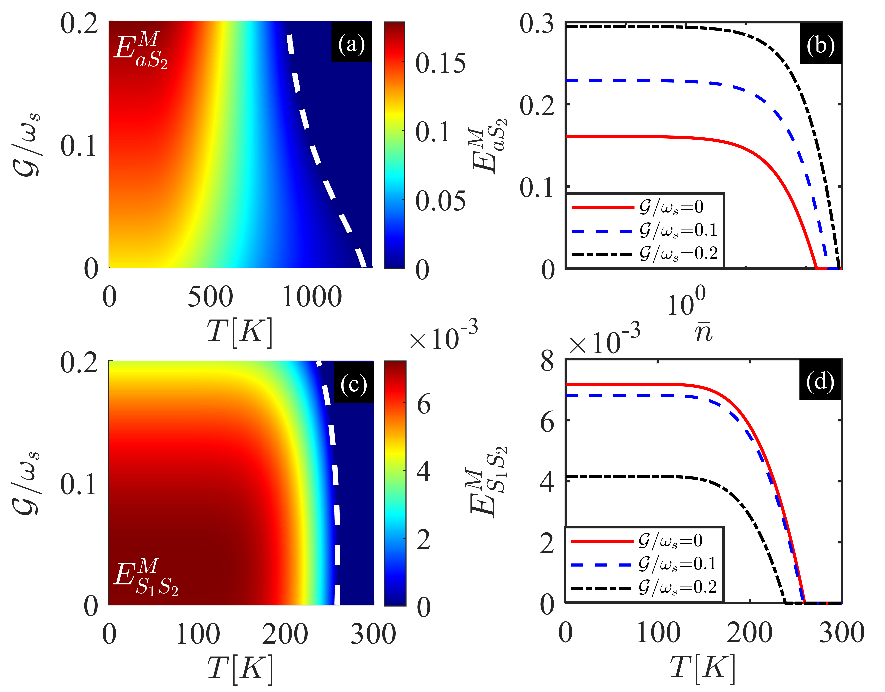}}
   \caption{Markovian logarithmic negativity in Eq.~\eqref{DSFX} for (a) optical-vibration entanglement $E_{aS_2}^{M}$ and (c,d) vibration-vibration entanglement $E_{S_1S_2}^{M}$ is plotted as a function of temperature $T$ and DOPA strength $\mathcal{G}$. (b) $E_{aS_2}^{M}$ is plotted as a function of thermal phonon number $\bar{N}_{th}$ for different DOPA strengths $\mathcal{G}$. The red solid, blue dashed, and black dash-dotted lines respectively correspond to $\mathcal{G} /\omega_s = 0, 0.1,$ and $0.2$. The white dashed lines denote critical temperature of entanglement with DOPA fixed for (a,c). The parameters chosen are $F/\omega_s = 16$ with $\Delta/\omega_s = 0.8$ for (a,b) and $F/\omega_s = 60$ with $\Delta/\omega_s = 2$ for (c,d). We take $\Lambda = 50$ for (a,c,d) and $\Lambda = 0$ for (b), while other parameters are the same as in Fig.~\ref{fig2abcd}.}\label{fig8abcd}
 \end{figure}

Temperature is an important factor for entanglement in open quantum systems because thermal noise leads to rapid decoherence and destruction of quantum entanglement. Figure~\ref{fig8abcd} shows robustness of bipartite entanglement against thermal noise under Markovian approximation.
In Fig.~\ref{fig8abcd}(a), optical-vibration entanglement $E_{aS_2}^M$ exhibits strong robustness existing at $T \approx 1000$ K. In Fig.~\ref{fig8abcd}(b), we plot $E_{aS_2}^M$ against thermal phonon number $\bar{N}_{th}$. The red solid, blue dashed, and black dash-dotted lines respectively correspond to ${\cal G} /{\omega _s}$ = 0, 0.1, and 0.2. We find that $E_{aS_2}^M$ can exist with a wide range of thermal noise from $\bar{N}_{th} = 0.01$ to $\bar{N}_{th} = 200$. This indicates that optical-vibration entanglement can emerge far above room temperature ($T \approx 300$ K), which is difficult to realize in conventional COM systems. This robustness stems from advantages of molecular COM systems: larger coupling, higher vibration frequencies, and stronger collective effects.
In Fig.~\ref{fig8abcd}(c) and (d), vibration-vibration entanglement $E_{S_1S_2}^M$ demonstrates strong robustness against thermal noise, despite being obtained via indirect cavity coupling. Although $E_{S_1S_2}^M$ decreases as $T$ rises, it can be maintained until at $T \approx 238$ K. The presence of DOPA affects magnitude for both types of entanglement but does not alter trend of thermal robustness.
From Fig.~\ref{fig8abcd}(a) and (c), we observe that both types of entanglement completely disappear once temperature exceeds a certain threshold (white dashed line) under Markovian approximation. This constraint leads to consider whether memory effects in non-Markovian reservoir can recover bipartite entanglement with DOPA fixed. We will discuss details in following sections.
\section{bipartite ENTANGLEMENT IN NON-MARKOVIAN reservoir}\label{four}
In this section, we investigate influences of non-Markovian effects on entanglement with DOPA fixed. For this purpose, we consider that cavity couples to non-Markovian reservoir consisting of a series of boson modes \cite{HPBreuer880210022016,IdeVega890150012017}. In rotating frame, the Hamiltonian~(\ref{HamiltonianEnd}) becomes
\begin{equation}
\begin{aligned}
\hat{H}_{NM} =& {\Delta _a}{\hat{a}^\dagger }\hat{a} + \sum\limits_{l = 1}^2 {{\omega _s}\hat{S}_l^\dagger {\hat{S}_l}}  + \sum\limits_{l = 1}^2 {{J_l}{\hat{a}^\dagger }\hat{a}(\hat{S}_l^\dagger  + {\hat{S}_l})}\\
&  + iF ({\hat{a}^\dagger } - \hat{a}) + i{\cal G} ({e^{i\theta }}{\hat{a}^{\dagger 2}} - {e^{ - i\theta }}{\hat{a}^2})\\
& + \sum\limits_k {{\Delta _k}\hat{c}_k^\dagger {\hat{c}_k} + i} \sum\limits_k {({W_k}\hat{a}} \hat{c}_k^\dagger  - W_k^ * {\hat{a}^\dagger }{\hat{c}_k}),
\label{HamiltonianNonMarkovian}
\end{aligned}
\end{equation}
where $\Delta_k = \omega_k - \omega_L$ defines detuning of $k$th mode (eigenfrequency $\omega_k$) of reservoir from driving field. $\hat{c}_k(\hat{c}_k^\dagger)$ is annihilation (creation) operator of reservoir. $W_k$ denotes coupling coefficient between cavity and reservoir. To obtain linearized form of Heisenberg-Langevin equations in Eqs.~(\ref{LangevinequationsNM1})-(\ref{LangevinequationsNM4}), we replace each operator with a steady-state amplitude plus a quantum fluctuation term $\hat{a}(t) = \alpha_a + \delta\hat{a}(t)$, $\hat{S}_1(t) = \beta_1 + \delta\hat{S}_1(t)$, $\hat{S}_2(t) = \beta_2 + \delta\hat{S}_2(t)$, and $\hat{c}_k(t) = c_k + \delta\hat{c}_k(t)$, where steady-state solutions in non-Markovian reservoir can be obtained from Eqs.~\eqref{mean_beta1}, \eqref{mean_beta2}, and \eqref{meanalphanew} (see Appendix \ref{Der12} for more details) as ${\beta _1} = { - i{J_1}{{\left| {{\alpha _a}} \right|}^2}}/{(i{\omega _s} + {\gamma _1})}$, ${\beta _2} = { - i{J_2}{{\left| {{\alpha _a}} \right|}^2}}/{(i{\omega _s} + {\gamma _2})}$, and ${\alpha _a} = F/[(i\Delta  + \kappa ) - 2{\cal G} {e^{i\theta }}]$. For quantum fluctuation operators, we have
\begin{align}
\delta\dot{\hat{a}}(t) =& -i\Delta \delta\hat{a}(t) - i\sum_{l=1}^{2} G_l [\delta\hat{S}_l^\dagger(t) + \delta\hat{S}_l(t)] \nonumber \\&+ 2{\cal G} e^{i\theta}\delta\hat{a}^\dagger(t) - \sum_k W_k^* \delta\hat{c}_k(t), \label{fluctuationa}\\
\delta\dot{\hat{S}}_1(t) =& -(i\omega_s + \gamma_1)\delta\hat{S}_1(t) - iG_1 [\delta\hat{a}^\dagger(t) + \delta\hat{a}(t)] \nonumber \\&+ \sqrt{2\gamma_1}\hat{S}_1^{in}(t), \label{fluctuationb1}\\
\delta\dot{\hat{S}}_2(t) =& -(i\omega_s + \gamma_2)\delta\hat{S}_2(t) - iG_2 [\delta\hat{a}^\dagger(t) + \delta\hat{a}(t)] \nonumber \\&+ \sqrt{2\gamma_2}\hat{S}_2^{in}(t), \label{fluctuationb2}\\
\delta\dot{\hat{c}}_k(t) =& -i\Delta_k \delta\hat{c}_k(t) + W_k \delta\hat{a}(t). \label{fluctuatione}
\end{align}
Substituting $\delta\hat{c}_k(t)$ solved by Eq.~(\ref{fluctuatione}) into Eq.~(\ref{fluctuationa}) gives
\begin{equation}
    \begin{aligned}
    \delta\dot{\hat{a}}(t) = &-i\Delta_a \delta\hat{a}(t) - i\sum_{l=1}^2 G_l [\delta\hat{S}_l^\dagger(t) + \delta\hat{S}_l(t)] \\
    &+ 2{\cal G} e^{i\theta}\delta\hat{a}^\dagger(t) + \delta\hat{Z}(t) - \int_0^t d\tau f(t-\tau)\delta\hat{a}(\tau),
    \end{aligned}
    \label{fluctuationintegrodiff}
\end{equation}
where $\delta\hat{Z}(t) = \int_{-\infty}^\infty d\tau h^*(t-\tau)\delta\hat{a}_{in}(\tau)$ denotes noise fluctuation operator. The input field operator is defined as $\delta\hat{a}_{in}(t) = \sum_k e^{-i\Delta_k t}\delta\hat{c}_k(0)/ \sqrt{2\pi}$. The definitions of impulse response function $h(t)$ and correlation function $f(t)$ can be found in Appendix \ref{Der12}. Different from mean-field case with $\langle \hat{Z}(t) \rangle = 0$, fluctuation operator $\delta\hat{Z}(t)$ represents non-zero colored noise source driving system.
To ensure Eqs.~(\ref{fluctuationa})-(\ref{fluctuatione}) being closed, we introduce auxiliary mode $\hat{C}(t)$ as $\sqrt{\kappa\lambda}\delta\hat{C}(t) \equiv  \delta\hat{Z}(t) - \int_0^t d\tau \kappa\lambda e^{-\lambda(t-\tau)}\delta\hat{a}(\tau)$ by considering Lorentzian spectral density (\ref{zz2e}), which leads to $\delta\dot{\hat{C}}(t) = -\lambda\delta\hat{C}(t) - \sqrt{\kappa\lambda}\delta\hat{a}(t) + \sqrt{2\lambda}\hat{a}_{in}(t)$ (see the derivation in Appendix~\ref{C}). In this case, Eqs.~(\ref{fluctuationa})-(\ref{fluctuatione}) become
\begin{align}
    \delta\dot{\hat{a}}(t) =& -i\Delta \delta\hat{a}(t) - i\sum_{l=1}^2 G_l [\delta\hat{S}_l^\dagger(t) + \delta\hat{S}_l(t)] \nonumber\\&+ 2{\cal G} e^{i\theta}\delta\hat{a}^\dagger(t) + \sqrt{\kappa\lambda}\delta\hat{C}(t), \label{NEWfluctuationa} \\
    \delta\dot{\hat{S}}_1(t) =& -(i\omega_s + \gamma_1)\delta\hat{S}_1(t) - iG_1[\delta\hat{a}(t) + \delta\hat{a}^\dagger(t)] \nonumber\\&+ \sqrt{2\gamma_1}\hat{S}_1^{in}(t), \label{NEWfluctuationb1} \\
    \delta\dot{\hat{S}}_2(t) =& -(i\omega_s + \gamma_2)\delta\hat{S}_2(t) - iG_2[\delta\hat{a}(t) + \delta\hat{a}^\dagger(t)] \nonumber\\&+ \sqrt{2\gamma_2}\hat{S}_2^{in}(t), \label{NEWfluctuationb2} \\
    \delta\dot{\hat{C}}(t) =& -\lambda\delta\hat{C}(t) - \sqrt{\kappa\lambda}\delta\hat{a}(t) + \sqrt{2\lambda}\hat{a}_{in}(t). \label{NEWfluctuatione}
\end{align}
In limit of infinite spectral width ($\lambda \to \infty$), the auxiliary mode $\hat{C}(t)$ can be eliminated, recovering Markovian linearized Langevin equations in Eqs.~(\ref{linearizedLangevinequations1})-(\ref{linearizedLangevinequations3}).
To analyze entanglement in non-Markovian reservoir, we introduce quadrature operators for auxiliary mode as $\delta \hat{X}_{C}=[\delta\hat{C}(t)+\delta\hat{C}^{\dagger}(t)]/\sqrt{2}$ and $\delta \hat{Y}_{C}=i[\delta\hat{C}^{\dagger}(t)-\delta\hat{C}(t)]/\sqrt{2}$ with corresponding input noise quadratures $\hat{X}_C^{in} = [\hat{a}_{in}(t) + \hat{a}_{in}^\dagger(t)]/\sqrt{2}$ and $\hat{Y}_C^{in} = i[\hat{a}_{in}^\dagger(t) - \hat{a}_{in}(t)]/\sqrt{2}$. Equations~(\ref{NEWfluctuationa})-(\ref{NEWfluctuatione}) can be written as $\dot{\hat{U}}^{NM}(t) = A^{NM}\hat{U}^{NM}(t) + \hat{B}^{NM}(t)$ with drift matrix
\begin{equation}
\label{NMPYJZ}
\resizebox{0.95\hsize}{!}{$
A^{NM} =
\begin{pmatrix}
2{\cal G} \cos \theta & 2{\cal G} \sin \theta + \Delta & 0 & 0 & 0 & 0 & \sqrt{\kappa\lambda} & 0 \\
2{\cal G} \sin \theta - \Delta & -2{\cal G} \cos \theta & -2G_1 & 0 & -2G_2 & 0 & 0 & \sqrt{\kappa\lambda} \\
0 & 0 & -\gamma_1 & \omega_s & 0 & 0 & 0 & 0 \\
-2G_1 & 0 & -\omega_s & -\gamma_1 & 0 & 0 & 0 & 0 \\
0 & 0 & 0 & 0 & -\gamma_2 & \omega_s & 0 & 0 \\
-2G_2 & 0 & 0 & 0 & -\omega_s & -\gamma_2 & 0 & 0 \\
-\sqrt{\kappa\lambda} & 0 & 0 & 0 & 0 & 0 & -\lambda & 0 \\
0 & -\sqrt{\kappa\lambda} & 0 & 0 & 0 & 0 & 0 & -\lambda
\end{pmatrix}
$},
\end{equation}
where extended vector of quadrature fluctuation operators $\hat{U}^{NM}(t) = (\delta\hat{X}_a, \allowbreak \delta\hat{Y}_a, \allowbreak \delta\hat{O}_1, \allowbreak \delta\hat{P}_1, \allowbreak \delta\hat{O}_2, \allowbreak \delta\hat{P}_2, \allowbreak \delta\hat{X}_C, \allowbreak \delta\hat{Y}_C)^{\mathrm{T}}$ and noise operators vector $\hat{B}^{NM}(t) = (0, 0, \allowbreak \sqrt{2\gamma_1} \hat{O}_1^{in}, \sqrt{2\gamma_1} \hat{P}_1^{in}, \allowbreak \sqrt{2\gamma_2} \hat{O}_2^{in}, \sqrt{2\gamma_2} \hat{P}_2^{in}, \allowbreak \sqrt{2\lambda} \hat{X}_C^{in}, \sqrt{2\lambda} \hat{Y}_C^{in})^{\mathrm{T}}$. The steady-state quantum fluctuations in non-Markovian reservoir are characterized by $8 \times 8$ covariance matrix $V^{NM}$ with matrix elements $V^{NM}_{ij} = \langle U_{i}^{NM}(\infty)U_{j}^{NM}(\infty) + U_{j}^{NM}(\infty)U_{i}^{NM}(\infty)\rangle / 2$ ($i, j = 1, 2, \cdots, 8$). The covariance matrix $V^{NM}$ is obtained by Lyapunov equation $A^{NM}V^{NM} + V^{NM}(A^{NM})^{\mathrm{T}} = -D^{NM}$, where $D^{NM} = \mathrm{diag} [0, 0, (2\bar{N}_1 + 1)\gamma_1, (2\bar{N}_1 + 1)\gamma_1, (2\bar{N}_2 + 1)\gamma_2, (2\bar{N}_2 + 1)\gamma_2, \lambda, \lambda]$ represents diffusion matrix with matrix elements $D_{ij}^{NM}$ defined by $\langle \hat{B}_i^{NM}(t)\hat{B}_j^{NM}(t') + \hat{B}_j^{NM}(t')\hat{B}_i^{NM}(t) \rangle/2 = D_{ij}^{NM} \delta(t-t')$.
To quantify bipartite entanglement between cavity and collective vibrational modes, we adopt non-Markovian logarithmic negativity $E^{NM}$ as measure with
\begin{equation}
    E^{NM} = \max [0, -\ln(2\xi^{NM})],\label{DSFXNM}
\end{equation}
where $\xi^{NM} = 2^{-1/2} \{ \oplus - [ \oplus^2 - 4\det( V_{sub}^{NM})]^{1/2} \}^{1/2}$ with $\oplus = \det(\eta_1) + \det(\eta_2) - 2\det(\eta_3)$. Since auxiliary mode $\hat{C}(t)$ is introduced to deal with non-Markovian effects, we need trace over its degrees of freedom to focus on system under study. In this case, we have ${V_{sub}^{NM}} = \left( \begin{smallmatrix} \eta_1 & \eta_3 \\ \eta_3^\mathrm{T} & \eta_2 \end{smallmatrix} \right)$, which is $4 \times 4$ sub-matrix extracted from whole $8 \times 8$ covariance matrix $V^{NM}$. $\eta_1, \eta_2$, and $\eta_3$ are $2 \times 2$ submatrices. Similar to Markovian approximation, we only show results for $E_{aS_2}^{NM}$ and $E_{S_1S_2}^{NM}$ as typical cases.
\begin{figure}[h]
   \centerline{
  \includegraphics[width=7.8cm, height=5.55cm, clip]{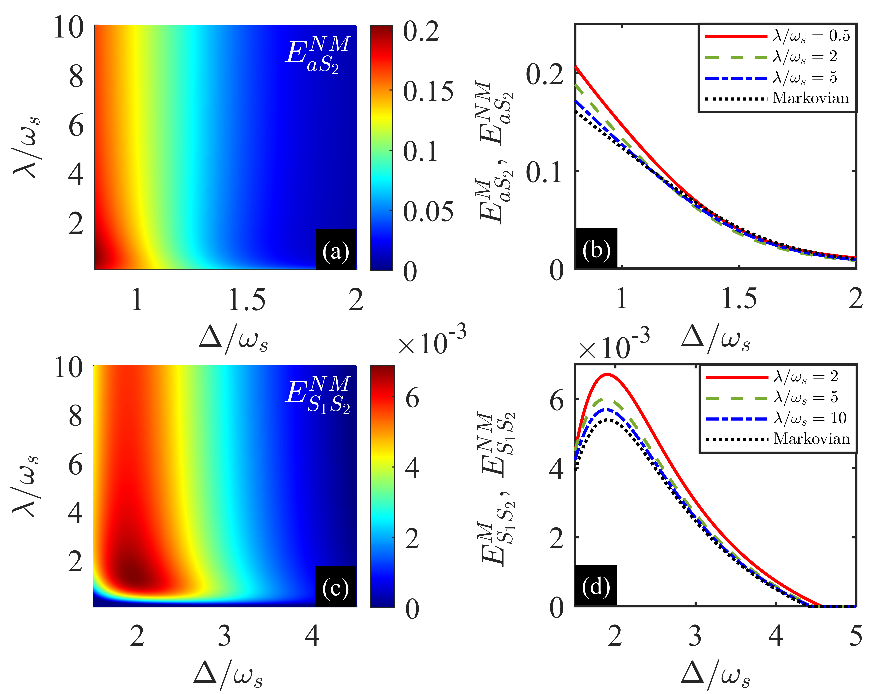}}
   \caption{Non-Markovian logarithmic negativity $E^{NM}$ in Eq.~\eqref{DSFXNM} is plotted as a function of detuning $\Delta$ for (a,b) optical-vibration entanglement $E_{aS_2}^{NM}$ and (c,d) vibration-vibration entanglement $E_{S_1S_2}^{NM}$ with different spectrum widths $\lambda$ without participation of DOPA (\({\cal G}= 0\)), while $E^{M}$ is given by Eq.~\eqref{DSFX} under Markovian approximation. The red solid, green dashed, and blue dash-dotted lines respectively correspond to $\lambda/\omega_s = 0.5$, $2$, and $5$ in (b) as well as $\lambda/\omega_s = 2$, $5$, and $10$ in (d), while black dotted lines denote Markovian approximation. The parameters chosen are $F/\omega_s = 16$ for (a,b) and $F/\omega_s = 60$ for (c,d), while other parameters are same as Fig.~{\ref{fig2abcd}}.}\label{fig9abcd}
 \end{figure}
\section{INFLUENCES OF NON-MARKOVIAN EFFECTS ON bipartite entanglement}\label{five}
Figure~\ref{fig9abcd} shows bipartite entanglement as a function of cavity detuning $\Delta$ and spectral width $\lambda$ in absence of DOPA (${\cal G} = 0$). The red solid, green dashed, and blue dash-dotted lines respectively correspond to $\lambda/\omega_s = 0.5$, $2$, and $5$ in (b) as well as $\lambda/\omega_s = 2$, $5$, and $10$ in (d), while black dotted lines denote Markovian approximation. The optical-vibration entanglement $E_{aS_2}^{NM}$ in Fig.~\ref{fig9abcd}(a) and (b) reaches maximum around $\Delta/\omega_s \approx 0.8$, whereas vibration-vibration entanglement $E_{S_1S_2}^{NM}$ in Fig.~\ref{fig9abcd}(c) and (d) reaches maximum around $\Delta/\omega_s \approx 2$. Specifically, both types of entanglement can be enhanced in the regime of small spectral width $\lambda$ with strong non-Markovian effects, which originate from excitation backflow obtained through interaction between cavity and non-Markovian reservoir. To be specific, decreasing spectral width $\lambda$ from Markovian approximation to $\lambda/\omega_s = 0.5$ for $E_{aS_2}^{NM}$ and $\lambda/\omega_s = 2$ for $E_{S_1S_2}^{NM}$, peak values increases from $0.1611$ to $0.2040$ and from $0.0054$ to $0.0069$, respectively.
 \begin{figure}[h]
   \centerline{
   \includegraphics[width=7.8cm, height=7.0cm, clip]{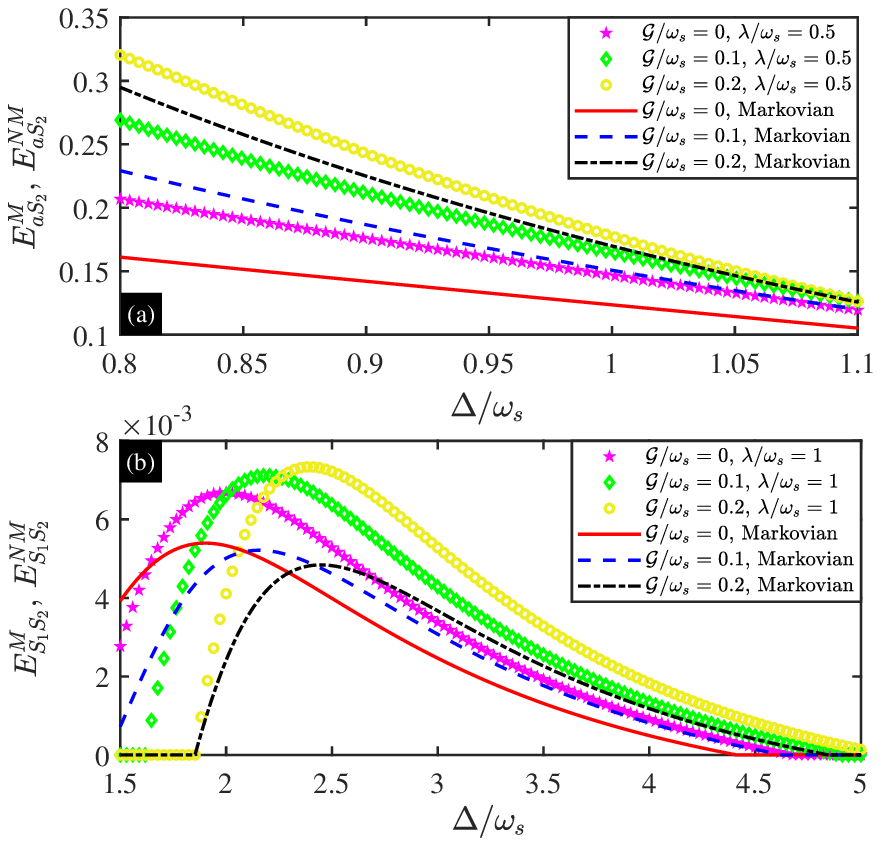}}
   \caption{Non-Markovian logarithmic negativity $E^{NM}$ in Eq.~\eqref{DSFXNM} is plotted as a function of detuning $\Delta$, while $E^{M}$ is given by Eq.~\eqref{DSFX} under Markovian approximation. The pink star, green diamond, and yellow circle denote that non-Markovian effects are considered for optical-vibration entanglement $E_{aS_2}^{NM}$ with $\lambda/\omega_s = 0.5$ in (a) and vibration-vibration entanglement $E_{S_1S_2}^{NM}$ with $\lambda/\omega_s = 1$ in (b). The red solid, blue dashed, and black dash-dotted lines respectively correspond to ${\cal G}/\omega_s = 0$, $0.1$, and $0.2$ in Markovian reservoir. The parameters chosen are $F/\omega_s = 16$ for (a) and $F/\omega_s = 60$ for (b), while other parameters are same as Fig.~{\ref{fig2abcd}}.}\label{fig10ab}
 \end{figure}

In Fig.~\ref{fig10ab}, we plot non-Markovian logarithmic negativity $E^{NM}$ in Eq.~\eqref{DSFXNM} as a function of detuning $\Delta$ with DOPA strength ${\cal G}$ fixed, while $E^{M}$ is given by Eq.~\eqref{DSFX} under Markovian approximation. In Fig.~\ref{fig10ab}(a), we find that optical-vibration entanglement $E_{aS_2}^{NM}$ increases with ${\cal G}$ for Markovian and non-Markovian cases. Moreover, non-Markovian effects enhance $E_{aS_2}^{NM}$ around $\Delta/\omega_s \approx 0.8$ with spectral width $\lambda/\omega_s = 0.5$. Compared to Markovian case with $\mathcal{G}/\omega_s = 0.2$ (black dash-dotted line), peak value of $E_{aS_2}^{NM}$ increases from 0.2949 to 0.3139. In Fig.~\ref{fig10ab}(b), vibration-vibration entanglement $E_{S_1S_2}^{NM}$ decreases as DOPA strength increases from ${\cal G}/\omega_s = 0$ (red solid line) to ${\cal G}/\omega_s = 0.2$ (black dash-dotted line), indicating suppression effect in Markovian regimes. However, we find that DOPA leads to enhancement for $E_{S_1S_2}^{NM}$ by choosing spectral width $\lambda/\omega_s = 1$ and detuning $\Delta/\omega_s \approx 2.4$ in non-Markovian regimes. In contrast to Markovian case with ${\cal G}/\omega_s = 0.2$ (black dash-dotted line), peak value of $E_{S_1S_2}^{NM}$ increases from 0.0048 to 0.0074. Therefore, we can realize stronger entanglement with non-Markovian effects, which change roles of DOPA from suppression to enhancement for vibration-vibration entanglement.
\begin{figure}[h]
   \centerline{
   \includegraphics[width=7.8cm, height=7.0cm, clip]{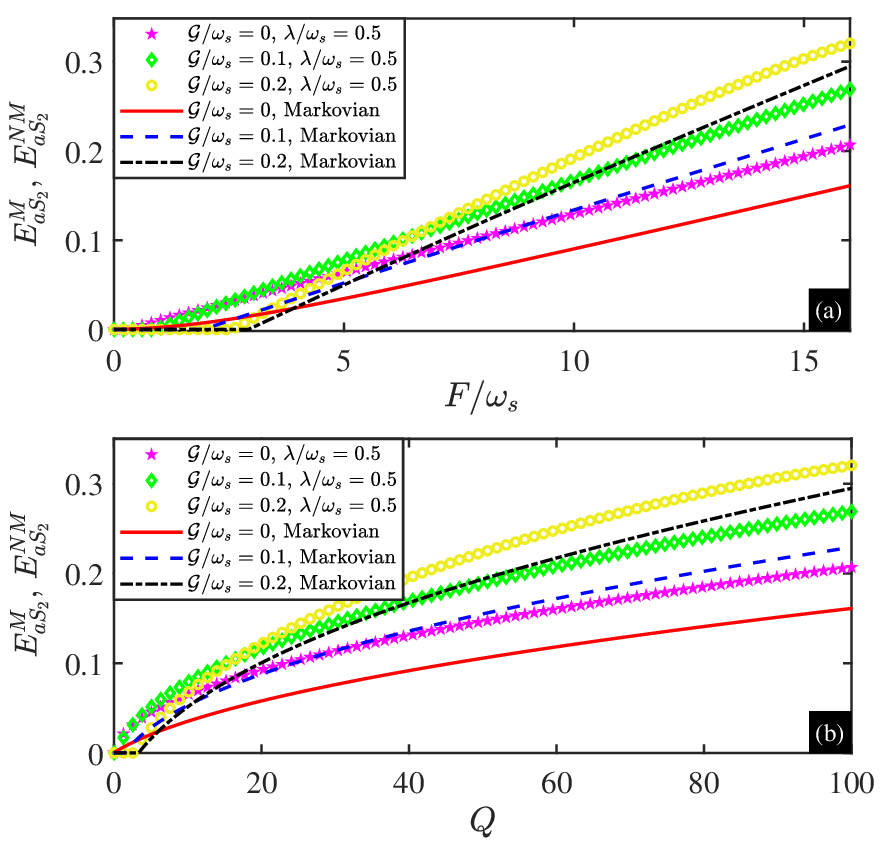}}
   \caption{Non-Markovian logarithmic negativity $E^{NM}$ in Eq.~\eqref{DSFXNM} for optical-vibration entanglement $E_{aS_2}^{NM}$ is plotted as a function of driving amplitude $F/\omega_s$ in (a) and molecular total number $Q$ in (b), while $E^{M}$ is given by Eq.~\eqref{DSFX} for $E_{aS_2}^{M}$ under Markovian approximation. The pink star, green diamond, and yellow circle denote that non-Markovian effects are considered for $E_{aS_2}^{NM}$ with $\lambda/\omega_s = 0.5$. The red solid, blue dashed, and black dash-dotted lines respectively correspond to ${\cal G}/\omega_s = 0$, $0.1$, and $0.2$ in Markovian reservoir. The parameters chosen are $\Delta/\omega_s = 0.8$ for (a,b) with $F/\omega_s = 16$ for (b), while other parameters are the same as in Fig.~\ref{fig2abcd}.}\label{fig11ab}
 \end{figure}

 In Fig.~\ref{fig11ab}, we plot optical-vibration entanglement $E_{aS_2}^M$ and $E_{aS_2}^{NM}$ as a function of driving amplitude $F$ in Fig.~\ref{fig11ab}(a) and total number of molecules $Q$ in Fig.~\ref{fig11ab}(b). We find that both $E_{aS_2}^M$ and $E_{aS_2}^{NM}$ increase with both $F$ and $Q$. Compared to Markovian reservoir, non-Markovian effects and DOPA further enhance collective effects and entanglement. To be specific, non-Markovian cases $E_{aS_2}^{NM}$ denoted by pink star (${\cal G}/\omega_s = 0$), green diamond (${\cal G}/\omega_s = 0.1$), and yellow circle (${\cal G}/\omega_s = 0.2$) with $\lambda/\omega_s = 0.5$ are stronger than corresponding Markovian results represented by red solid (${\cal G}/\omega_s = 0$), blue dashed (${\cal G}/\omega_s = 0.1$), and black dash-dotted (${\cal G}/\omega_s = 0.2$) lines.
 \begin{figure}[h]
   \centerline{
   \includegraphics[width=7.0cm, height=2.6cm, clip]{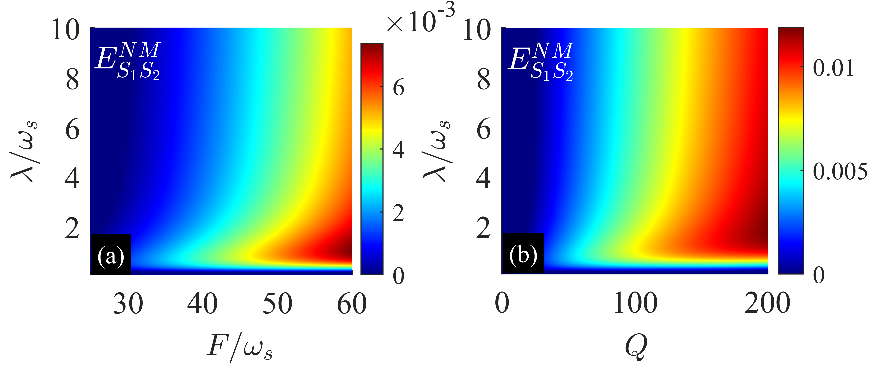}}
   \caption{Non-Markovian logarithmic negativity $E^{NM}$ in Eq.~\eqref{DSFXNM} for vibration-vibration entanglement $E_{S_1S_2}^{NM}$ is plotted as a function of spectrum width $\lambda/\omega_s$ and (a) driving amplitude $F$, and (b) molecular total number $Q$. The parameters chosen are $\mathcal{G}/\omega_s = 0.2$ and $\Delta/\omega_s = 2.4$ for (a,b) with $F/\omega_s = 60$ and $\Lambda = Q/2$ for (b), while other parameters are the same as in Fig.~\ref{fig2abcd}.}\label{fig12ab}
 \end{figure}
 Figure~\ref{fig12ab} shows vibration-vibration entanglement $E_{S_1S_2}^{NM}$ as a function of spectral width $\lambda$ and driving amplitude $F$ in Fig.~\ref{fig12ab}(a), and molecular total number $Q$ in Fig.~\ref{fig12ab}(b) with DOPA strength fixed ${\cal G}/\omega_s = 0.2$. We find that $E_{S_1S_2}^{NM}$ can be enhanced as $F$ and $Q$ increase. Moreover, $E_{S_1S_2}^{NM}$ is stronger in region of small spectral width $\lambda$. These results further verify that non-Markovian effects can make DOPA enhance vibration-vibration entanglement. By choosing strong driving $F/\omega_s = 60$ and $\Delta/\omega_s = 2.4$, the combination of collective effects with $Q=200$ and DOPA strength ${\cal G}/\omega_s = 0.2$ leads to maximum value $0.0119$ in non-Markovian regimes.

 \begin{figure}[h]
   \centerline{
   \includegraphics[width=7.8cm, height=7.0cm, clip]{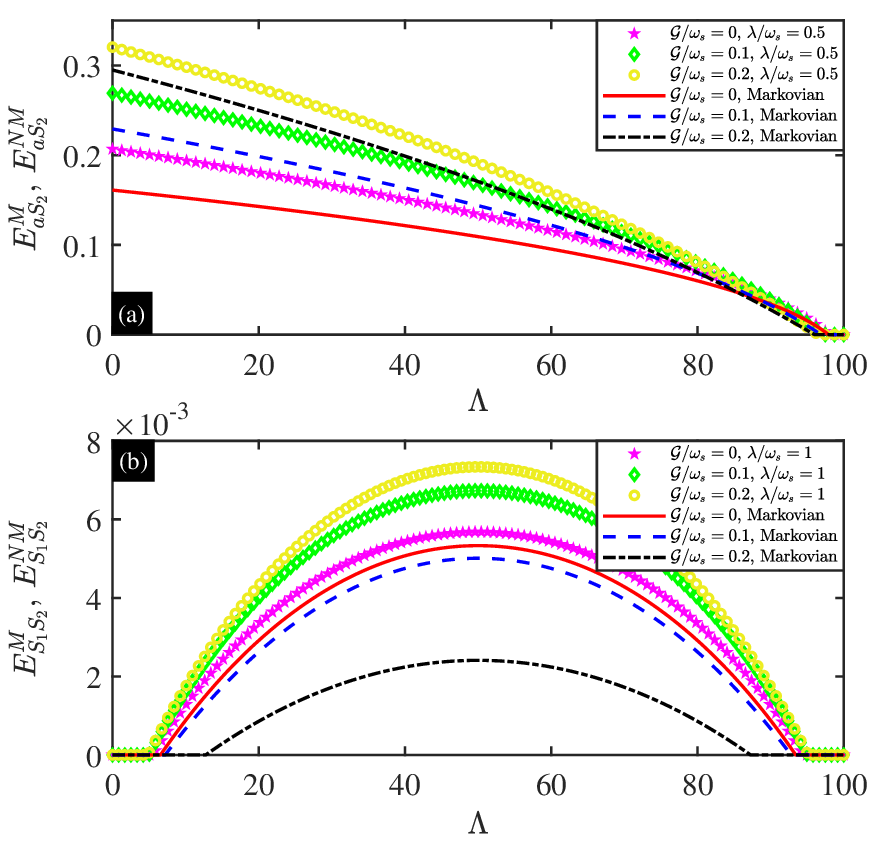}}
   \caption{Non-Markovian logarithmic negativity $E^{NM}$ in Eq.~\eqref{DSFXNM} is plotted as a function of molecular collective mode distribution $\Lambda$, while $E^{M}$ is given by Eq.~\eqref{DSFX} under Markovian approximation. The pink star, green diamond, and yellow circle denote that non-Markovian effects are considered for optical-vibration entanglement $E_{aS_2}^{NM}$ with $\lambda/\omega_s = 0.5$ and $\Delta/\omega_s = 0.8$ in (a), and vibration-vibration entanglement $E_{S_1S_2}^{NM}$ with $\lambda/\omega_s = 1$ and $\Delta/\omega_s = 2.4$ in (b). The red solid, blue dashed, and black dash-dotted lines respectively correspond to ${\cal G}/\omega_s = 0$, $0.1$, and $0.2$ in Markovian reservoir with $\Delta/\omega_s = 0.8$ for (a) and $\Delta/\omega_s = 2$ for (b). The parameters chosen are \(F/\omega_s = 16\) for (a) and \(F/\omega_s = 60\) for (b), while other parameters are same as Fig.~{\ref{fig2abcd}}.}\label{fig13ab}
 \end{figure}
Figure~\ref{fig13ab} plots non-Markovian logarithmic negativity $E^{NM}$ in Eq.~\eqref{DSFXNM} as a function of molecular distribution number $\Lambda$ with DOPA fixed, while $E^{M}$ is given by Eq.~\eqref{DSFX} under Markovian approximation. In Fig.~\ref{fig13ab}(a), optical-vibration entanglement $E_{aS_2}^{NM}$ decreases as $\Lambda$ increases. Moreover, $E_{aS_2}^{NM}$ increases across whole range of $\Lambda$ due to DOPA and non-Markovian effects. In Fig.~\ref{fig13ab}(b), vibration-vibration entanglement $E_{S_1S_2}^{NM}$ increases first and then decreases. The maximum $E_{S_1S_2}^{NM}$ is obtained when molecules are distributed between two collective vibrational modes ($\Lambda = Q/2$). In contrast to Markovian cases where $E_{S_1S_2}^{M}$ is minimized at DOPA strength ${\cal G}/\omega_s = 0.2$ (black dash-dotted line), non-Markovian effects can turn this suppression into enhancement, causing $E_{S_1S_2}^{NM}$ with ${\cal G}/\omega_s = 0.2$ (yellow circle) to reach maximum.
  \begin{figure}[h]
   \centerline{
   \includegraphics[width=7.8cm, height=7.0cm, clip]{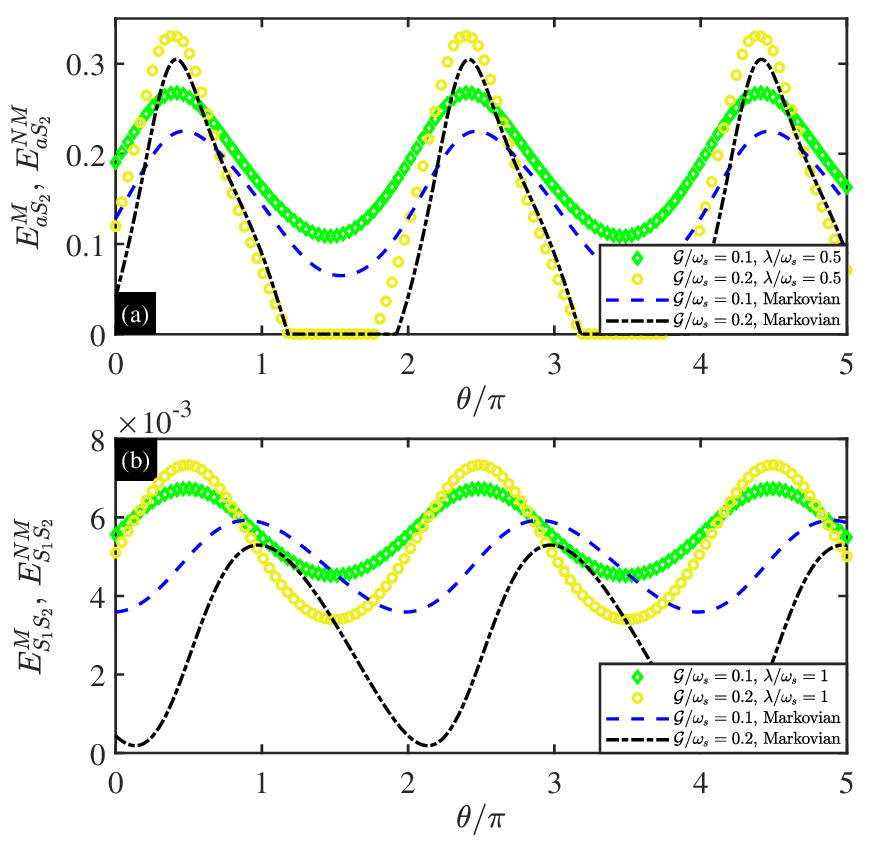}}
   \caption{Non-Markovian logarithmic negativity $E^{NM}$ in Eq.~\eqref{DSFXNM} is plotted as a function of DOPA phase $\theta$, while $E^{M}$ is given by Eq.~\eqref{DSFX} under Markovian approximation. The green diamond and yellow circle denote that non-Markovian effects are considered for optical-vibration entanglement $E_{aS_2}^{NM}$ with $\lambda/\omega_s = 0.5$ and $\Delta/\omega_s = 0.81$ in (a), and vibration-vibration entanglement $E_{S_1S_2}^{NM}$ with $\lambda/\omega_s = 1$ and $\Delta/\omega_s = 2.4$ in (b). The blue dashed and black dash-dotted lines respectively correspond to ${\cal G}/\omega_s = 0.1$ and $0.2$ in Markovian reservoir with $\Delta/\omega_s = 0.81$ for (a) and $\Delta/\omega_s = 2$ for (b). The parameters chosen are $F/\omega_s = 16$ for (a) and $F/\omega_s = 60$ for (b), while other parameters are same as Fig.~{\ref{fig2abcd}}.}\label{fig14ab}
 \end{figure}
\begin{figure}[h]
   \centerline{
   \includegraphics[width=7.8cm, height=7.0cm, clip]{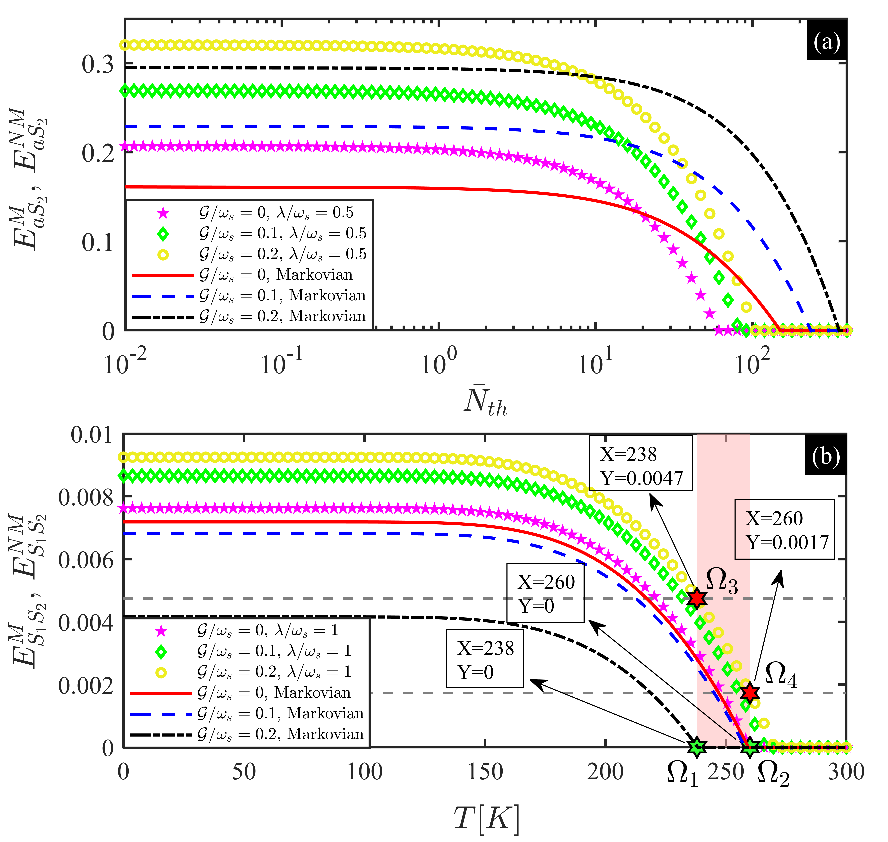}}
   \caption{Non-Markovian logarithmic negativity $E^{NM}$ in Eq.~\eqref{DSFXNM} is plotted as a function of thermal phonon number $\bar{N}_{th}$ in (a) and temperature $T$ in (b), while $E^{M}$ is given by Eq.~\eqref{DSFX} under Markovian approximation. The pink star, green diamond, and yellow circle denote that non-Markovian effects are considered for optical-vibration entanglement $E_{aS_2}^{NM}$ with $\lambda/\omega_s = 0.5$ and $\Delta/\omega_s = 0.8$ in (a), and vibration-vibration entanglement $E_{S_1S_2}^{NM}$ with $\lambda/\omega_s = 1$ and $\Delta/\omega_s = 2.4$ in (b). The red solid, blue dashed, and black dash-dotted lines respectively correspond to ${\cal G}/\omega_s = 0$, $0.1$, and $0.2$ in Markovian reservoir with $\Delta/\omega_s = 0.8$ for (a) and $\Delta/\omega_s = 2$ for (b). The green stars $\Omega_1$ and $\Omega_2$ indicate the disappearance of entanglement. The red stars $\Omega_3$ and $\Omega_4$ indicate the regeneration of entanglement. Two gray dashed lines represent non-zero entanglement values. The parameters chosen are \(F/\omega_s = 16\) for (a) and \(F/\omega_s = 60\) for (b), while other parameters are same as Fig.~{\ref{fig2abcd}}.}\label{fig15ab}
 \end{figure}
\begin{figure}[h]
   \centerline{
   \includegraphics[width=7.8cm, height=7.0cm, clip]{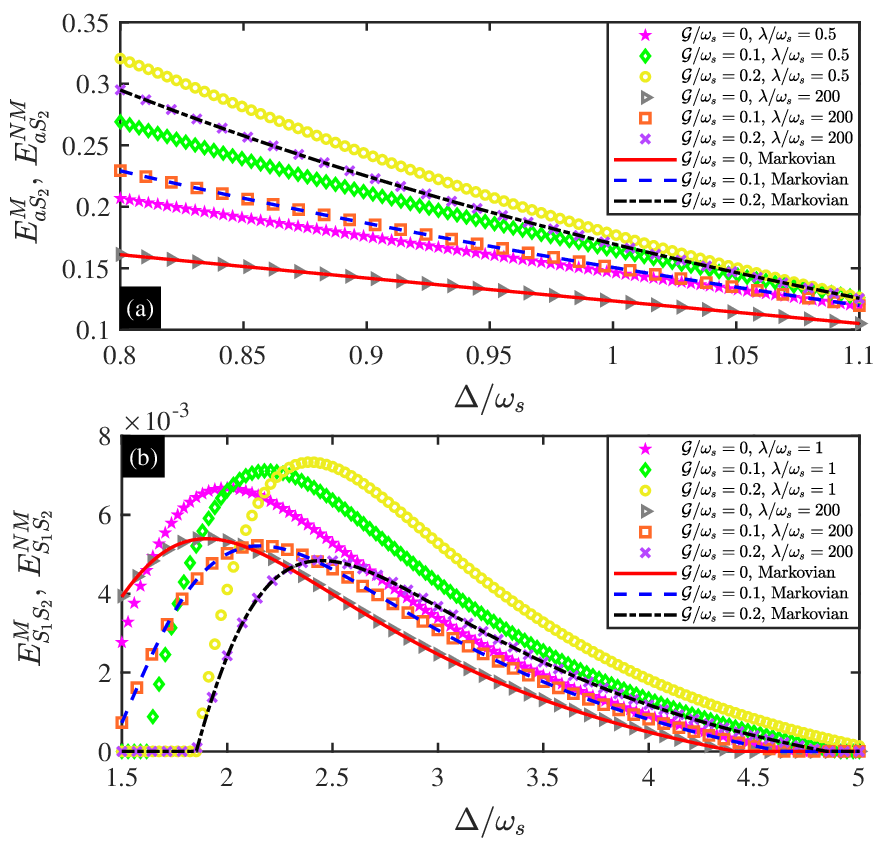}}
   \caption{Non-Markovian logarithmic negativity $E^{NM}$ in Eq.~\eqref{DSFXNM} is plotted as a function of detuning $\Delta$, while $E^{M}$ is given by Eq.~\eqref{DSFX} under Markovian approximation. The pink star, green diamond, and yellow circle denote that non-Markovian effects are considered for optical-vibration entanglement $E_{aS_2}^{NM}$ with $\lambda/\omega_s = 0.5$ in (a) and vibration-vibration entanglement $E_{S_1S_2}^{NM}$ with $\lambda/\omega_s = 1$ in (b). The gray triangle, orange square, and purple cross denote corresponding non-Markovian results with large spectral width $\lambda/\omega_s = 200$. The red solid, blue dashed, and black dash-dotted lines respectively correspond to ${\cal G}/\omega_s = 0$, $0.1$, and $0.2$ in Markovian reservoir. The parameters chosen are \(F/\omega_s = 16\) for (a) and \(F/\omega_s = 60\) for (b), while other parameters are same as Fig.~{\ref{fig2abcd}}.}\label{fig16ab}
 \end{figure}

Figure~\ref{fig14ab} shows non-Markovian logarithmic negativity $E^{NM}$ in Eq.~\eqref{DSFXNM} as a function of DOPA phase $\theta$, while $E^{M}$ is given by Eq.~\eqref{DSFX} under Markovian approximation. In Fig.~\ref{fig14ab}(a), $E_{aS_2}^{NM}$ reaches peak values around $\theta \approx (2q + 1/2)\pi$ (i.e., $0.5\pi$, $2.5\pi$, and $4.5\pi$) and decreases to minima around $\theta \approx (2q + 3/2)\pi$ (i.e., $1.5\pi$ and $3.5\pi$). In Fig.~\ref{fig14ab}(b), we find that $E_{S_1S_2}^{NM}$ increases with non-Markovian effects and DOPA fixed. In particular, phase values for $E_{S_1S_2}^{NM}$ become consistent with $E_{aS_2}^{NM}$, reaching peak values around $\theta \approx 0.5\pi$, $2.5\pi$, and $4.5\pi$. From this, we can maximize both types of entanglement by adjusting same phase values.

Figure~\ref{fig15ab} demonstrates robustness of bipartite entanglement against thermal noise with DOPA and non-Markovian effects. In Fig.~\ref{fig15ab}(a), optical-vibration entanglement $E_{aS_2}^{NM}$ exhibits strong robustness against thermal noise. Specifically, non-Markovian effects and DOPA can enhance $E_{aS_2}^{NM}$ within the range of thermal phonon numbers from $\bar{N}_{th}=0.01$ to $10$ compared to Markovian approximation. In Fig.~\ref{fig15ab}(b), we show vibration-vibration entanglement $E_{S_1S_2}^{NM}$ as a function of temperature $T$, focusing on cases with DOPA strength ${\cal G}/\omega_s = 0.2$ (black dash-dotted line and yellow circle). However, once $T$ increases to the range of $238$~K to $260$~K, $E_{S_1S_2}^{NM}$ can not be maintained and disappear (green stars $\Omega_1$ and $\Omega_2$) in Markovian reservoir. Nevertheless, $E_{S_1S_2}^{NM}$ can be revived within this temperature range by tuning spectral width $\lambda$ in non-Markovian reservoir, exhibiting regeneration of entanglement. The regeneration points $\Omega_3$ and $\Omega_4$ are marked by red stars, while two gray dashed lines denote non-zero entanglement values. $E_{S_1S_2}^{NM}$ regenerates from 0 with $\Omega_1$ to $0.0047$ with $\Omega_3$ at $T=238$~K. Similarly, $E_{S_1S_2}^{NM}$ increases from 0 with $\Omega_2$ to $0.0017$ with $\Omega_4$ at $T=260$~K. This means that non-Markovian reservoir plays important roles in regenerating vibration-vibration entanglement at high temperatures.

Moreover, we simulate system under condition of large spectral width $\lambda$. In Fig.~\ref{fig16ab}(a) and (b), bipartite entanglement exhibits differences between Markovian and non-Markovian cases with spectrum widths $\lambda/\omega_s = 0.5$ and $1$, respectively. However, system dynamics changes as spectrum width $\lambda$ increases. We separately plot non-Markovian limit with large spectral width $\lambda/\omega_s = 200$ (gray triangle, orange square, and purple cross) and Markovian approximation (red solid, blue dashed, and black dash-dotted lines). This figure shows consistency of bipartite entanglement between non-Markovian limit and Markovian approximation. It stems from the fact that correlation function $f(t)$ approaches $2\kappa\delta(t)$ when $\lambda$ is close to infinity, which causes Eqs.~\eqref{meanalphanew} and \eqref{fluctuationintegrodiff} in non-Markovian reservoir returning to Eqs.~\eqref{mean1} and \eqref{linearizedLangevinequations1} under Markovian approximation.

\section{conclusion}\label{six}
In summary, we have proposed a DOPA mediated scheme to manipulate bipartite entanglement in molecular COM system. By comparing results under Markovian and non-Markovian reservoirs, we revealed different mechanisms of DOPA and non-Markovian effects on entanglement.
Under Markovian approximation, DOPA can enhance optical-vibration entanglement and suppress that between vibrational modes. The constraint has been overcome by considering non-Markovian effects, which not only can increase optical-vibration entanglement but reverse roles of DOPA on vibration-vibration entanglement from suppression to enhancement (regeneration of entanglement) with DOPA fixed. Although vibration-vibration entanglement suffers from sudden disappearance in Markovian regimes within temperature range of 238 K to 260 K with other parameters fixed, non-Markovian effects can revive it. Due to collective enhancement effects, optical-vibration entanglement exhibits remarkable robustness against thermal noise and can exist at room temperature with experimentally feasible parameters.

\section*{AUTHOR CONTRIBUTIONS}
This project was conceived by C. Shang and H. Z. Shen. The methodology was developed by Wenyao Hu, H. N. Liu, X. C. Zhang, C. Shang, Yan-Hui Zhou, and H. Z. Shen. The project was supervised by C. Shang, Yan-Hui Zhou, and H. Z. Shen. The original draft of the manuscript was written by Wenyao Hu and H. Z. Shen. The manuscript was reviewed and edited by Wenyao Hu, H. N. Liu, X. C. Zhang, C. Shang, Yan-Hui Zhou, and H. Z. Shen.

\section*{ACKNOWLEDGMENTS}
This work was supported by Science and Technology Development Plan Project of Jilin Province (Grant No.~20250102007JC), National Natural Science Foundation of China (NSFC) under Grants No.~12274064 and No. 12374333, and the Hakubi Projects of RIKEN.

\section*{CONFLICTS OF INTEREST}
The authors declare no conflicts of interest.

\section*{DATA AVAILABILITY STATEMENT}
The data that support the findings of this study are available from the corresponding author upon reasonable request.

\section*{CODE AVAILABILITY STATEMENT}
Code used to generate data in this study are available from the corresponding author upon reasonable request.\\

\noindent \textbf{Correspondence} and requests for materials should be addressed to C. Shang or H. Z. Shen.

\appendix
\section{Heisenberg-Langevin equations and linearization under Markovian approximation}
\label{Der1}
The total Hamiltonian of system reads
\begin{equation}
\begin{aligned}
\hat{H}(t)= &\omega_{a}\hat{a}^{\dagger}\hat{a}+\sum_{n=1}^{Q}\omega_s\hat{s}_{n}^{\dagger}\hat{s}_{n}+\sum_{n=1}^{Q}J_s\hat{a}^{\dagger}\hat{a}(\hat{s}_{n}^{\dagger}+\hat{s}_{n})\\
&+iF(\hat{a}^{\dagger}e^{-i\omega_{L}t}-\hat{a}e^{i\omega_{L}t})\\
&+i\mathcal{G}(e^{i\theta}\hat{a}^{\dagger2}{e^{ - i{\omega _p}t}}-e^{-i\theta}\hat{a}^{2}{e^{i{\omega _p}t}}).
\label{HamiltonianTotal}
\end{aligned}
\end{equation}
In rotating frame at frequency ${{\omega _L}}$ and set ${\omega _p} = 2{\omega _L}$, the transformed Hamiltonian is given by
\begin{equation}
\begin{aligned}
\hat{H}=&\Delta_{a}\hat{a}^{\dagger}\hat{a}+\sum_{n=1}^{Q}\omega_s\hat{s}_{n}^{\dagger}\hat{s}_{n}+\sum_{n=1}^{Q}J_s\hat{a}^{\dagger}\hat{a}(\hat{s}_{n}^{\dagger}+\hat{s}_{n})\\
&+iF(\hat{a}^{\dagger}-\hat{a})+i{\cal G}(e^{i\theta}\hat{a}^{\dagger 2}-e^{-i\theta}\hat{a}^{2}).
\label{HamiltonianRevlove}
\end{aligned}
\end{equation}
Setting molecules collective operators defined below Eq.~(\ref{HamiltonianEnd}), Eq.~(\ref{HamiltonianRevlove}) becomes Eq.~(\ref{HamiltonianEnd}).
By introducing loss terms, Heisenberg-Langevin equations with Eq.~(\ref{HamiltonianEnd}) give
\begin{align}
\frac{d\hat{a}}{dt}=&-(i\Delta_{a}+\kappa)\hat{a}-iJ_{1}\hat{a}(\hat{S}_{1}^{\dagger}+\hat{S}_{1})-iJ_{2}\hat{a}(\hat{S}_{2}^{\dagger}+\hat{S}_{2}) \nonumber \\
&+2{\cal G} e^{i\theta}\hat{a}^{\dagger}+F+\sqrt{2\kappa}\hat{a}^{in}, \label{Langevinequations1} \\
\frac{d\hat{S}_{1}}{dt}=&-(i\omega_s+\gamma_{1})\hat{S}_{1}-iJ_{1}\hat{a}^{\dagger}\hat{a}+\sqrt{2\gamma_{1}}\hat{S}_{1}^{in}, \label{Langevinequations2} \\
\frac{d\hat{S}_{2}}{dt}=&-(i\omega_s+\gamma_{2})\hat{S}_{2}-iJ_{2}\hat{a}^{\dagger}\hat{a}+\sqrt{2\gamma_{2}}\hat{S}_{2}^{in}, \label{Langevinequations3}
\end{align}
where ${\kappa}$ and $\gamma_{l=1,2}$ denote decay rates of cavity and collective vibrational modes, respectively. $\hat{\mathcal{T}}^{in}$ ($\hat{\mathcal{T}}$ = $\hat{a}$, $\hat{S}_{1}$, $\hat{S}_{2}$) denotes input noise operators. For collective vibrational modes, corresponding noise operators are defined as $\hat{S}_{1}^{in}=\sum_{n=1}^{\Lambda}\hat{s}_{n}^{in}/\sqrt{\Lambda}$ and $\hat{S}_{2}^{in}=\sum_{n=\Lambda+1}^{Q}\hat{s}_{n}^{in}/\sqrt{Q-\Lambda}$, which are assumed to have zero mean values and satisfy
$\langle \hat{a}^{in}(t)\hat{a}^{in\dagger}(t^{\prime})\rangle =\delta(t-t^{\prime})$, $\langle \hat{S}_{l}^{in}(t)\hat{S}_{l}^{in\dagger}(t^{\prime})\rangle =(\bar{N}_{l}+1)\delta(t-t^{\prime})$, and $\langle \hat{S}_{l}^{in\dagger}(t)\hat{S}_{l}^{in}(t^{\prime})\rangle =\bar{N}_{l}\delta(t-t^{\prime})$ \cite{CWGardiner2000} with $\bar{N}_l = \{\exp(\omega_s / k_B T) - 1\}^{-1}$ being thermal phonon number at temperature $T$. ${{k_B}}$ is Boltzmann constant. Since both collective modes ${S_1}$ and ${S_2}$ share same frequency ${\omega _s} $, their thermal phonon numbers are equal with ${{\bar N}_1} = {{\bar N}_2} = \bar{N}_{th}$.
We linearize Eqs.~(\ref{Langevinequations1})-(\ref{Langevinequations3}) by expressing each operator as a sum of steady-state values and quantum fluctuation operators $\hat{a} = \alpha_a + \delta\hat{a}$, $\hat{S}_1 = \beta_1 + \delta\hat{S}_1$, and $\hat{S}_2 = \beta_2 + \delta\hat{S}_2$ with
\begin{align}
{{\dot \alpha }_a} =&  - (i{\Delta _a} + \kappa ){\alpha _a} - i{J_1}{\alpha _a}(\beta _1^ *  + {\beta _1}) - i{J_2}{\alpha _a}(\beta _2^ *  + {\beta _2}) \nonumber \\
&+ 2{\cal G} {e^{i\theta }}\alpha _a^ *  + F , \label{mean1} \\
{{\dot \beta }_1} =&  - (i{\omega _s} + {\gamma _1}){\beta _1} - i{J_1}{\left| {{\alpha _a}} \right|^2}, \label{mean2} \\
{{\dot \beta }_2} =&  - (i{\omega _s} + {\gamma _2}){\beta _2} - i{J_2}{\left| {{\alpha _a}} \right|^2}. \label{mean3}
\end{align}
The linearized quantum Langevin equations for fluctuation operators can be obtained as
\begin{align}
\delta\dot{\hat{a}} =& -(i\Delta+\kappa)\delta\hat{a}-iG_{1}(\delta\hat{S}_{1}^{\dagger}+\delta\hat{S}_{1})-iG_{2}(\delta\hat{S}_{2}^{\dagger}+\delta\hat{S}_{2}) \nonumber \\
&+ 2{\cal G} e^{i\theta}\delta\hat{a}^{\dagger}+\sqrt{2\kappa}\hat{a}^{in}, \label{linearizedLangevinequations1} \\
\delta\dot{\hat{S}}_{1} =& -(i\omega_s+\gamma_{1})\delta\hat{S}_{1}-iG_{1}(\delta\hat{a}+\delta\hat{a}^{\dagger})+\sqrt{2\gamma_{1}}\hat{S}_{1}^{in}, \label{linearizedLangevinequations2} \\
\delta\dot{\hat{S}}_{2} =& -(i\omega_s+\gamma_{2})\delta\hat{S}_{2}-iG_{2}(\delta\hat{a}+\delta\hat{a}^{\dagger})+\sqrt{2\gamma_{2}}\hat{S}_{2}^{in}, \label{linearizedLangevinequations3}
\end{align}
where we have defined linearized coupling strengths ${G_1} = \sqrt \Lambda {G_s}$ and ${G_2} = \sqrt {Q - \Lambda} {G_s}$ with $G_s = J_s \left| \alpha_a \right|$.

\section{Heisenberg-Langevin equations and linearization in non-Markovian reservoir}
\label{Der12}
The Heisenberg equations in Eq.~(\ref{HamiltonianNonMarkovian}) lead to
\begin{align}
\frac{d\hat{a}(t)}{dt} =& -i\Delta_a \hat{a}(t) - i\sum_{l=1}^{2} J_l \hat{a}(t)[\hat{S}_l^\dagger(t) + \hat{S}_l(t)] \nonumber \\
& + 2{\cal G} e^{i\theta}\hat{a}^\dagger(t) + F - \sum_k W_k^* \hat{c}_k(t), \label{LangevinequationsNM1} \\
\frac{d\hat{S}_1(t)}{dt} =& -(i\omega_s + \gamma_1)\hat{S}_1(t) - iJ_1\hat{a}^\dagger(t)\hat{a}(t) + \sqrt{2\gamma_1}\hat{S}_1^{in}, \label{LangevinequationsNM2} \\
\frac{d\hat{S}_2(t)}{dt} =& -(i\omega_s + \gamma_2)\hat{S}_2(t) - iJ_2\hat{a}^\dagger(t)\hat{a}(t) + \sqrt{2\gamma_2}\hat{S}_2^{in}, \label{LangevinequationsNM3} \\
\frac{d\hat{c}_k(t)}{dt} =& -i\Delta_k\hat{c}_k(t) + W_k\hat{a}(t). \label{LangevinequationsNM4}
\end{align}
Under conditions of strong driving, we linearize Eqs.~(\ref{LangevinequationsNM1})-(\ref{LangevinequationsNM4}) by each operator as a sum of steady-state mean value and small fluctuation defined below Eq.~(\ref{HamiltonianNonMarkovian}) with
\begin{align}
{{\dot \alpha }_a} =& -i{\Delta _a} {\alpha _a} - i\sum_{l=1}^{2} {J_l}{\alpha _a}(\beta _l^ *  + {\beta _l}) \nonumber + 2{\cal G} {e^{i\theta }}\alpha _a^ *  \\
&+ F - \sum_k W_k^* {c_k}, \label{mean_alpha} \\
{{\dot \beta }_1} =&  - (i{\omega _s} + {\gamma _1}){\beta _1} - i{J_1}{\left| {{\alpha _a}} \right|^2}, \label{mean_beta1} \\
{{\dot \beta }_2} =&  - (i{\omega _s} + {\gamma _2}){\beta _2} - i{J_2}{\left| {{\alpha _a}} \right|^2}, \label{mean_beta2} \\
{{\dot c }_k} =& -i\Delta_k c_k + W_k {\alpha _a}. \label{mean_ck}
\end{align}
Substituting $c_k(t)$ obtained by Eq.~(\ref{mean_ck}) into Eq.~(\ref{mean_alpha}), we get the non-Markovian integro-differential equation
\begin{equation}
\begin{aligned}
{{\dot \alpha }_a} =& -i{\Delta _a} {\alpha _a} - i\sum_{l=1}^{2} {J_l}{\alpha _a}(\beta _l^ *  + {\beta _l}) + 2{\cal G} {e^{i\theta }}\alpha _a^ *  \\
&+ F + Z(t) - \int_{0}^{t} d\tau \alpha_a(\tau) f(t - \tau),
\label{meanalphanew}
\end{aligned}
\end{equation}
where $Z(t) = \langle \hat{Z}(t) \rangle$ represents expectation value of reservoir operator defined as $\hat{Z}(t) = -\sum_k W_k^* \hat{c}_k(0) e^{-i\Delta_k t} = \int_{-\infty}^\infty d\tau h^*(t-\tau)\hat{a}_{in}(\tau)$ with input-field operator $\hat{a}_{in}(t) = \sum_k e^{-i\Delta_k t} \hat{c}_k(0) / \sqrt{2\pi}$ and impulse response function $h(t) = -\sum_k e^{i\Delta_k t} W_k/ \sqrt{2\pi}$, or $h(t-\tau) = (-1 / \sqrt{2\pi}) \int_{-\infty}^\infty e^{i(\omega'-\omega_L)(t-\tau)} W(\omega') d\omega'$ with the replacement $W_k \rightarrow W(\omega')$. $\hat{a}_{in}(t)$ is input field with zero means $a_{in}(t) = \langle \hat{a}_{in}(t) \rangle = 0$ for reservoir initialization in vacuum state, which leads to $Z(t) = \langle \hat{Z}(t) \rangle = 0$ and $\hat{a}_{in}(t)=\delta\hat{a}_{in}(t)$. The correlation function of reservoir is given by
$f(t) = \sum_{k} |W_k|^2 e^{-i\Delta_k t} = \int_{-\infty}^\infty \rho(\omega') e^{-i(\omega' - \omega_L)t} d\omega'$, where $\rho(\omega') = \sum_k |W_k|^2\delta(\omega' - \omega_k)$ represents spectral density of reservoir. $f(t)$ characterizes memory effects of reservoir acting on system.
We assume spectral response function as $W(\omega')=-\sqrt{\kappa/\pi}\lambda/[\lambda + i(\omega'-\omega_{L})]$, where $\lambda$ is spectrum width, while $\kappa$ denotes dissipation. The Lorentzian spectral density takes \cite{HPBreuer2002,HZShen1070537052023,HZShen1050237072022,HZShen950121562017,JFYang1090537122024,WZhang1090337012024,LDiosi850341012012} \begin{equation}
\begin{aligned}
\rho(\omega') = \kappa \lambda^2 / \{\pi [\lambda^2 + (\omega' - \omega_L)^2]\},
\label{zz2e}
\end{aligned}
\end{equation}
which can be realized in all-optical setups \cite{SJXiong1000321012019,FFFanchini1122104022014} and pseudomode methods \cite{HZShene715352026,DTamascelli1200304022018,ZXMan900621042014,LMazzola800121042009,
GPleasance960621052017}. With above $W(\omega')$ and $\rho(\omega')$, we get $h(t-\tau)=\sqrt{2\kappa}\lambda e^{-\lambda(t-\tau)}\Theta(t-\tau)$ and $f(t-\tau)=\kappa\lambda e^{-\lambda|t-\tau|}$, where $\Theta(t-t')$ is unit step function: $\Theta(t-t')=1$ for $t \geqslant t'$, while $\Theta(t-t')=0$ for $t < t'$, which represents a Gaussian Ornstein-Uhlenbeck process \cite{JJing1052404032010}. In wideband limit with $\lambda$ tending to infinity, the reservoir becomes memoryless, where $\rho(\omega') \to \kappa/\pi$ and $W(\omega') \to -\sqrt{\kappa/\pi}$. According to these, we have $f(t) = 2\kappa\delta(t)$ and $h(t) = \sqrt{2\kappa}\delta(t)$.

\vspace{2em}

\section{Derivation of auxiliary mode}\label{C}
\noindent To derive
\begin{equation}
\label{eqC1}
\delta\dot{\hat{C}}(t) = -\lambda\delta\hat{C}(t) - \sqrt{\kappa\lambda}\delta\hat{a}(t) + \sqrt{2\lambda}\hat{a}_{in}(t),
\end{equation}
above Eq.~(\ref{NEWfluctuationa}), we start with the definition of auxiliary mode 
\begin{equation}
\label{eq:aux_def}
\sqrt{\kappa\lambda}\delta\hat{C}(t) \equiv \delta\hat{Z}(t) - \int_0^t f(t-\tau)\delta\hat{a}(\tau)d\tau,
\end{equation}
where the correlation function is $f(t-\tau) = \kappa\lambda e^{-\lambda(t-\tau)}$. Taking the time derivative on both sides of Eq.~(\ref{eq:aux_def}) requires calculating the derivatives of noise fluctuation operator $\delta\hat{Z}(t)$ and memory integral term $\int_0^t f(t-\tau)\delta\hat{a}(\tau)d\tau$. 
For the noise fluctuation operator $\delta\hat{Z}(t) = \int_{-\infty}^{+\infty} d\tau h^*(t-\tau) \delta\hat{a}_{in}(\tau)$ with the impulse response function $h(t-\tau)=\sqrt{2\kappa}\lambda e^{-\lambda(t-\tau)}\Theta(t-\tau)$ and $\dot{h}^*(t-\tau) = -\lambda h^*(t-\tau) + \lambda\sqrt{2\kappa}\delta(t-\tau)e^{-\lambda(t-\tau)}$ due to $\dot{\Theta}(t) = \delta(t)$, we get
\begin{equation}
\label{eq:noise_deriv}
\delta\dot{\hat{Z}}(t) = -\lambda \delta\hat{Z}(t) + \lambda\sqrt{2\kappa}\delta\hat{a}_{in}(t).
\end{equation}
Differentiating $\int_0^t f(t-\tau) \delta\hat{a}(\tau) d\tau$ in Eq.~(\ref{eq:aux_def}) gives
\begin{small}
\begin{equation}
\label{eq:integral_deriv}
\frac{d}{dt} \left[ \int_0^t f(t-\tau) \delta\hat{a}(\tau) d\tau \right] = \kappa\lambda\delta\hat{a}(t) - \lambda \int_0^t f(t-\tau) \delta\hat{a}(\tau) d\tau.
\end{equation}
\end{small}Substituting Eqs.~(\ref{eq:noise_deriv}) and (\ref{eq:integral_deriv}) into  Eq.~(\ref{eq:aux_def}), we obtain
\begin{equation}
\label{eq:combine_deriv}
\begin{split}
\sqrt{\kappa\lambda}\delta\dot{\hat{C}}(t) =& -\lambda [ \delta\hat{Z}(t) - \int_0^t f(t-\tau) \delta\hat{a}(\tau) d\tau ] \\
& - \kappa\lambda\delta\hat{a}(t) + \lambda\sqrt{2\kappa}\delta\hat{a}_{in}(t).
\end{split}
\end{equation}
Substituting Eq.~(\ref{eq:aux_def}) into Eq.~(\ref{eq:combine_deriv}) and considering the input-field operator $\hat{a}_{in}(t)$ with zero means for reservoir initialization in vacuum state (leading to $\delta\hat{a}_{in}(t) = \hat{a}_{in}(t)$ due to $\hat{a}_{in}(t) = \langle\hat{a}_{in}(t)\rangle + \delta\hat{a}_{in}(t)$ and $\langle\hat{a}_{in}(t)\rangle = 0$), we obtain Eq.~(\ref{eqC1}).

\end{document}